\documentstyle[aps,eqsecnum,epsf]{revtex}

\begin{document}

\draft

\title{Statistical Mechanics of Vacancy and Interstitial Strings in Hexagonal Columnar Crystals}
\author{Shilpa Jain and David R. Nelson}
\address{Lyman Laboratory of Physics, Harvard University, Cambridge, Massachusetts 02138}
\maketitle

\begin{abstract}
Columnar crystals contain defects in the form of vacancy/interstitial loops or 
strings of vacancies and interstitials bounded by column ``heads'' and
``tails''. These 
defect strings are oriented by the columnar lattice and can change size and 
shape by movement of the ends and forming kinks along the length. Hence an 
analysis in terms of directed living polymers~\cite{Safran} is appropriate to 
study their 
size and shape distribution, volume fraction, etc. If the entropy of transverse 
fluctuations overcomes the string line tension in the crystalline phase, a 
string proliferation transition occurs, leading to a supersolid 
phase~\cite{Frey:defect}. We 
estimate the wandering entropy and examine the behaviour in the transition 
regime. We also calculate numerically the line tension of various species 
of vacancies and interstitials in a triangular lattice for power-law potentials 
as well as for a modified Bessel function interaction between columns as occurs 
in the case of flux lines in type-II superconductors or long 
polyelectrolytes in an ionic solution. We find that the centered 
\underline{interstitial} 
is the lowest energy defect for a very wide range of interactions; the 
symmetric vacancy is preferred only for extremely short interaction ranges.
\end{abstract}


\section{Introduction}
The physics of columnar crystals is relevant to 
the Abrikosov lattice of flux lines in Type-II superconductors and liquid 
crystalline materials like 
concentrated phases of long polymers or discotics. The stability of the 
columnar crystal has been investigated, and various mechanisms 
proposed for its melting. Conventional melting, which arises when phonon 
displacements reach a fixed fraction of the lattice constant, can easily be 
located via the Lindemann 
criterion~\cite{Nelson:directed,Jain:ice}. Melting destroys the two-dimensional 
crystalline order 
perpendicular to the columns leading to a nematic liquid of lines or columns, 
which is entangled at sufficiently high densities.

Crystal defects play an important role above the melting 
transition. If edge dislocations in the crystal proliferate, they drive the 
shear modulus to zero, leading to a liquid-like shear viscosity. 
However, dislocations alone cannot destroy the six-fold orientational order of 
the triangular 
lattice in a two-dimensional cross-section. Thus, provided disclination lines 
do not also proliferate, the resulting liquid of lines is 
hexatic, not isotropic~\cite{MarchNel:hexatic}. The screw component of the 
unbound dislocations leads to entanglement.
A finite concentration of unbound disclinations superimposed on
the hexatic liquid leads to isotropic in-plane order.  

Another kind of transition is brought about by vacancy/interstitial line 
defects in columnar crystals composed of long, continuous lines. As discussed 
in Ref.~\cite{Frey:defect}, under 
suitable conditions (such as high field and small interlayer coupling in 
layered superconductors), it can become favourable for these line defects to 
proliferate. If this happens at a temperature $T_d$ below the melting 
temperature $T_m$, then the phase that exists between $T_d$ and $T_m$ will be 
simultaneously crystalline and highly entangled. In the boson analogy of an 
aligned 
system of lines, where the lines represent two-dimensional bosons traveling in 
the ``time-like'' axial ($\hat{\mathbf z}$) direction~\cite{Nelson:directed}, 
such a phase is 
analogous to the supersolid phase of the bosonic system which incorporates 
vacancies and interstitials in its ground state. This entangled solid melts 
into an 
entangled liquid or an entangled hexatic at even higher temperatures.

The proliferation of vacancy or interstitial strings could also affect a 
crystal-to-hexatic transition mediated by dislocations. Dislocations in the 
columnar crystalline 
geometry are normally constrained to lie in the vertical plane formed by 
their Burger's vector and the $\hat{\mathbf z}$-axis, because a dislocation in 
a two-dimensional cross-section 
can move along the columnar axis only through glide parallel to its Burger's 
vector. Transverse motion (climb) would require it to absorb or emit vacancies 
or 
interstitials. This becomes possible in the supersolid phase, thus allowing 
dislocation loops to take on arbitrary non-planar configurations which would 
have to be included in the treatment of Ref.~\cite{MarchNel:hexatic} to 
study melting out of a supersolid phase~\cite{MR:supersolid}.

Vacancy/interstitial strings in a columnar crystal tend to be lines themselves 
because 
of the continuity of the columns. If the columns are constrained to be
continuous across the entire sample (as is the case for vortex lines in Type II 
superconductors), 
these defects must either thread the entire sample (Fig.~\ref{string}) or 
appear in vacancy/interstitial pairs forming 
loops (Fig.~\ref{loop})~\cite{Frey:defect}. The situation is different, 
however, for finite-length polymers, or columns of discotic liquid crystal 
molecules which can break and reform freely. 
As illustrated in Fig.~\ref{polymer}a, a slice through a low temperature 
configuration in a polymer columnar crystal (with translational order 
perpendicular to the column axis but not parallel to it) would consist of 
tightly bound polymer ``heads and tails''. At higher temperatures, however, the 
heads and tails will separate, either moving apart to form a vacancy string or 
sliding past each other to form a line of interstitials
(Fig.~\ref{polymer}b)~\cite{polymer-defects}. In columnar discotic crystals with
similar translational order, ``heads'' and ``tails'' are absent at low 
temperatures, but appear spontaneously when vacancy and interstitial strings 
are excited (Fig.~\ref{discotic}). (Head and tail defects appear superficially 
like dislocations in the cross sections shown in
Figs.~\ref{polymer} and~\ref{discotic}.
A three-dimensional analysis of lines and columns in 
neighbouring sheets like that shown in Figs.~\ref{string} and~\ref{loop} is 
necessary to clearly reveal that these are strings of vacancies and 
interstitials.)

\noindent
\begin{minipage}{3.4in}
\begin{figure}
\centering
\leavevmode
\epsfxsize=3.2in
\epsfbox{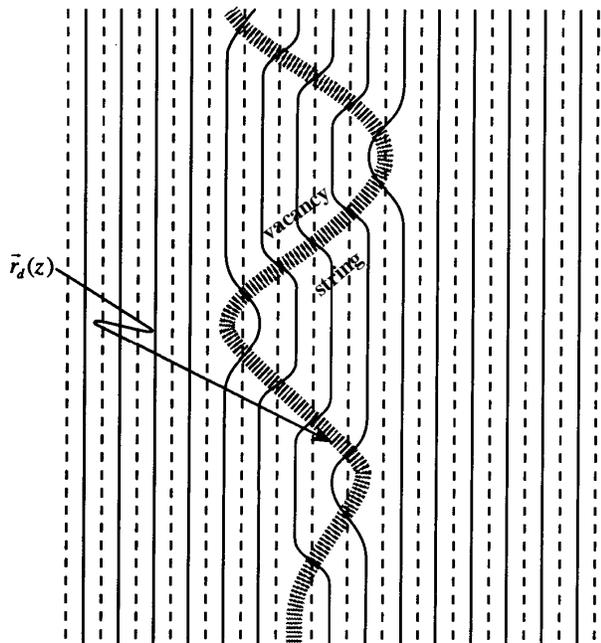}
\caption{Vacancy string ${\mathbf r}_d(z)$ (thick dashed curve) meandering 
through a columnar crystal. Dashed lines represent columns just above or below 
the plane of the figure. (Taken from Ref.~\protect\cite{Frey:defect}.)}
\label{string}
\end{figure}
\end{minipage}
\hfill
\begin{minipage}{3.4in}
\begin{figure}
\centering
\leavevmode
\epsfxsize=2.8in
\epsfbox{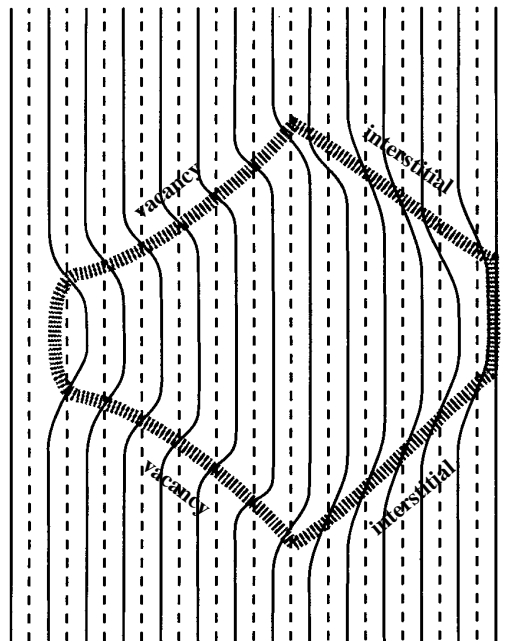}
\smallskip
\caption{Vacancy-interstitial loop in a columnar crystal. Dashed lines 
represent columns just above or below the plane of the figure. (Taken from
Ref.~\protect\cite{Frey:defect}.)}
\label{loop}
\end{figure}
\end{minipage}
\medskip

\noindent
\begin{minipage}{3.4in}
\begin{figure}
\centering
\leavevmode
\epsfxsize=3in
\epsfbox{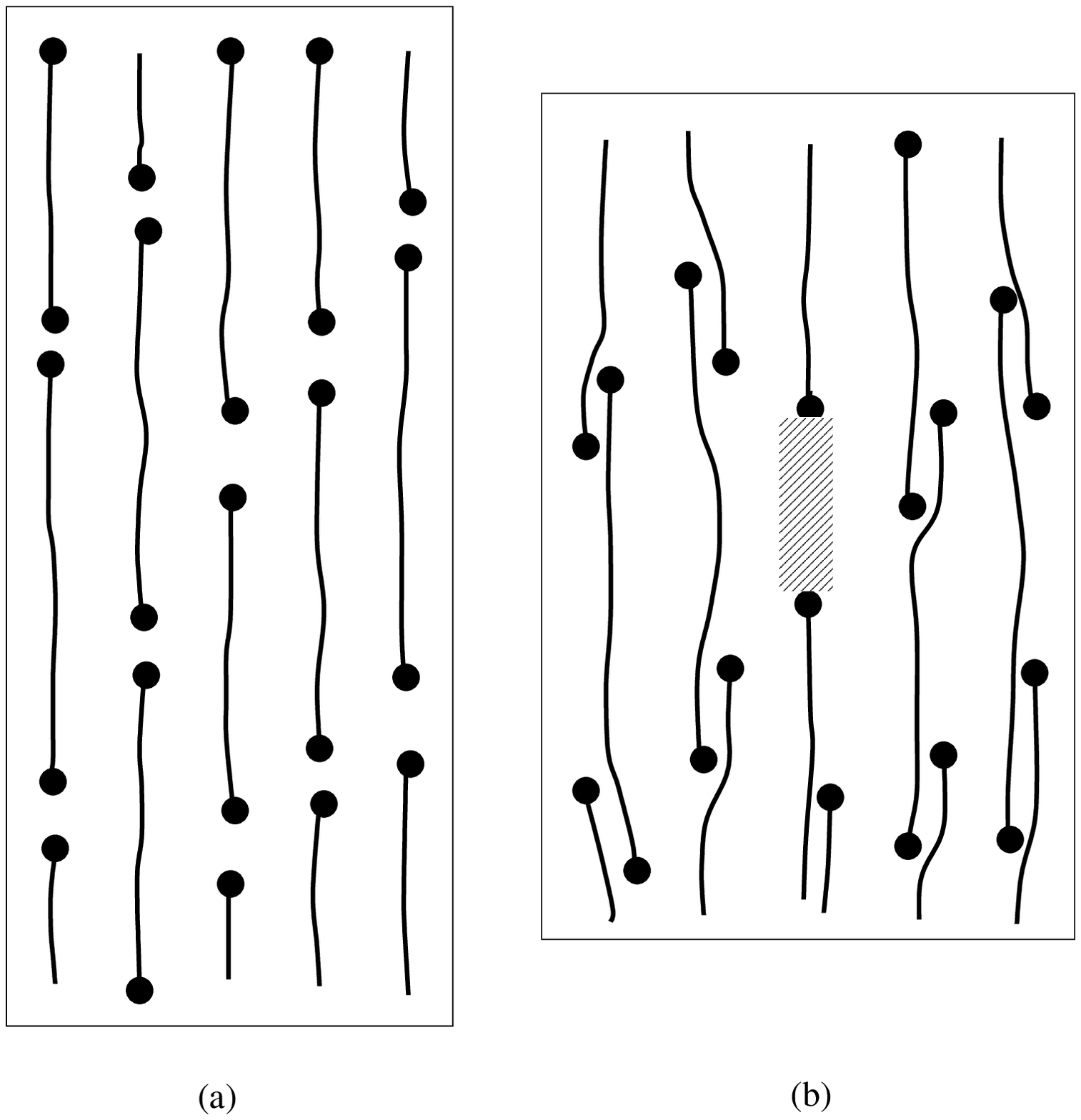}
\caption{Formation of vacancy/interstitial strings by sliding of polymers 
within columns in a columnar crystal of finite-length polymers.}
\label{polymer}
\end{figure}
\end{minipage}
\hfill
\begin{minipage}{3.4in}
\begin{figure}
\centering
\leavevmode
\epsfxsize=3in
\epsfbox{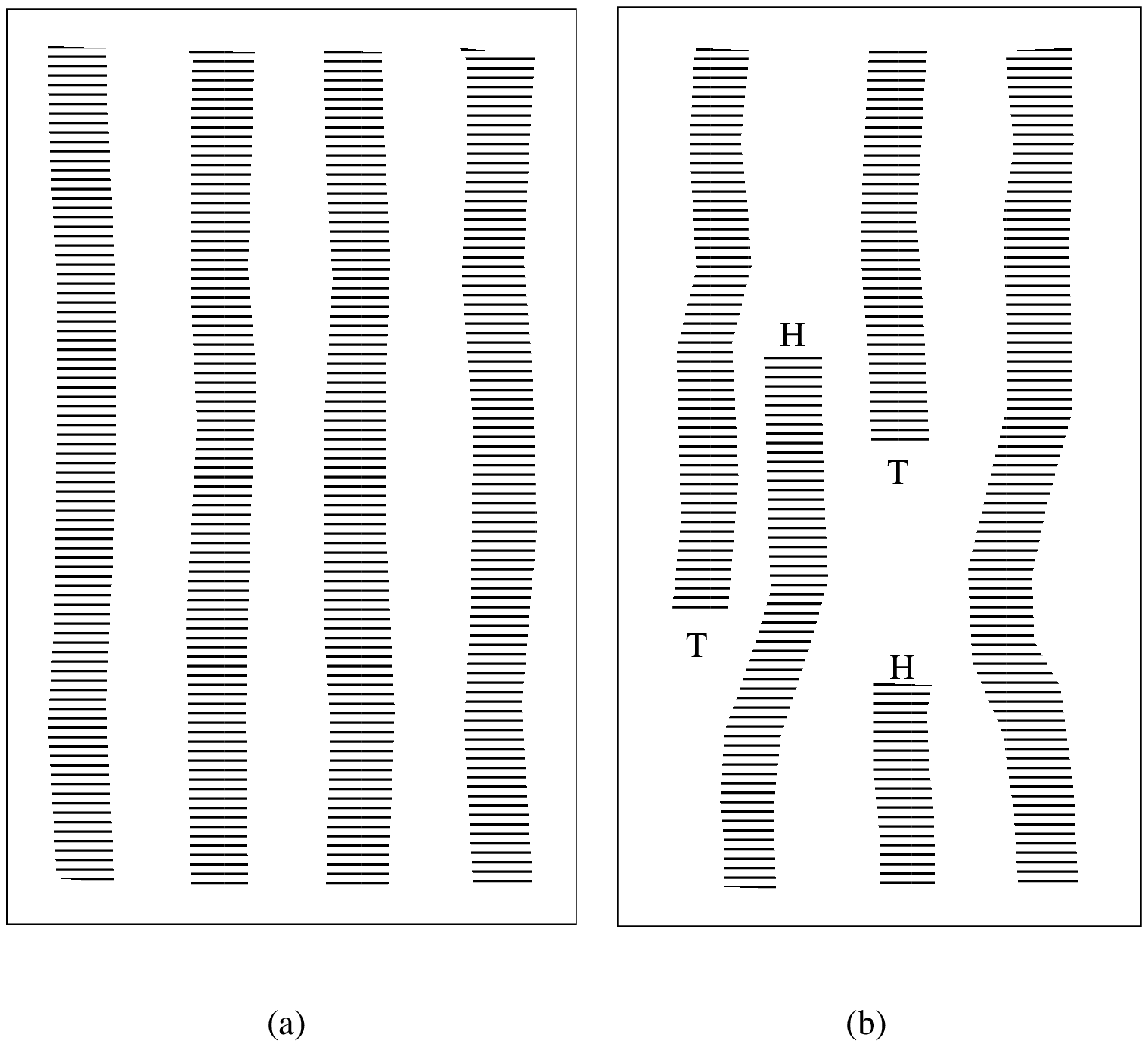}
\bigskip
\caption{Formation of vacancy/interstitial strings by sliding of polymers 
within columns in a columnar crystal of finite-length polymers.}
\label{discotic}
\end{figure}
\end{minipage}
\medskip

Unlike dislocation lines, these strings (and loops) are not 
constrained to be planar: the lines can jump to any neighbouring lattice 
site as they traverse the crystal. Several horizontal jumps connecting a head 
to a tail are shown in Fig.~\ref{jumps}. Note that \underline{left}ward 
deflections of the interstitial segment connecting a head to a tail are 
accompanied by \underline{right}ward deflections of the lines or columns 
themselves. A typical string can be approximated 
by an alternating sequence of straight segments and kinks joining the head of 
one column or polymer chain to the tail of another (see Fig.~\ref{walk}).
Vacancy/interstitial strings are suppressed at low temperatures because they 
have a finite line tension, and hence an energy proportional to their length. 
At higher temperatures, heads and tails can move apart, forming variable-length 
strings that wander or ``diffuse'' perpendicular to their length by forming 
kinks. These strings thus resemble living polymers~\cite{Safran}, except that 
they are directed, on average, along the $\hat{\mathbf z}$-axis. In polymer 
crystals, the number of such strings is determined by the fixed concentration 
of heads and tails. In columnar discotic crystals, heads and tails can be 
created freely, and it is appropriate to treat their statistical mechanics in a 
grand canonical ensemble by introducing a head/tail fugacity, similar to the 
fugacity which controls defect concentrations in theories of vortex or 
dislocation unbinding transitions~\cite{Nelson:trans}. We assume here that we 
can treat polymer crystals using the same formalism provided we tune the 
head/tail fugacity to achieve the fixed concentration determined by the mean 
polymer length. Long polymers imply a dilute distribution of heads and tails. 
We exclude, for simplicity, the possibility of hairpin excitations in polymer 
systems, which can be regarded as doubly quantized interstitial excitations 
leading to a higher energy. As we shall see, the sharp defect proliferation 
transition discussed in Ref.~\cite{Frey:defect} is blurred when there is a 
finite concentration of heads and tails in equilibrium.

\noindent
\begin{minipage}{3.4in}
\begin{figure}
\centering
\leavevmode
\epsfxsize=2in
\epsfbox{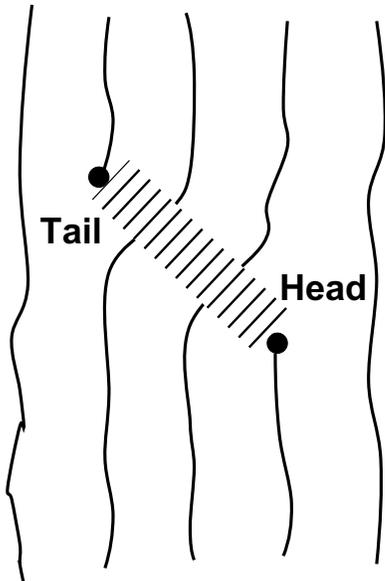}
\caption{Illustration of a vacancy string (thick dashed curve) joining a column 
head to another column's tail in a columnar crystal composed of long-chain 
polymers.}
\label{jumps}
\end{figure}
\end{minipage}
\hfill
\begin{minipage}{3.4in}
\begin{figure}
\centering
\leavevmode
\epsfxsize=2.8in
\epsfbox{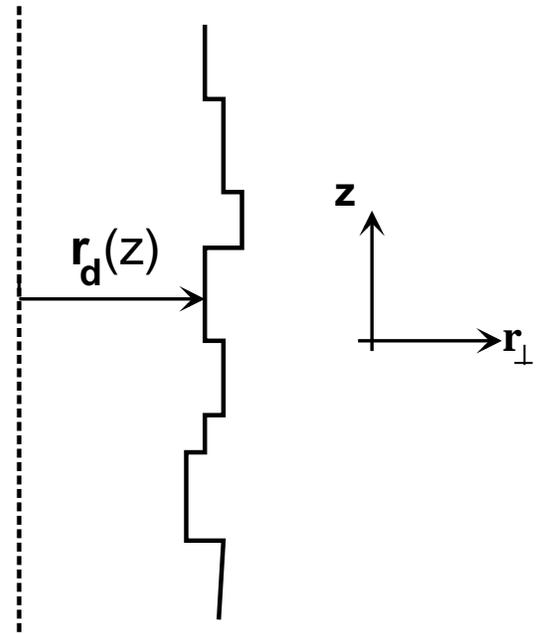}
\caption{Schematic of a defect string (composed of straight segments and kinks) 
wandering through the columnar crystal.}
\label{walk}
\end{figure}
\end{minipage}
\medskip

Given an appropriate combination of parameters, namely, low line tension 
combined with 
head/tail and kink energies comparable to the temperature, the entropy of 
diffusion of the strings can overcome the line tension and lead to 
string proliferation, allowing heads and tails to separate to arbitrarily large 
distances. As in its bosonic counterpart, there exists 
off-diagonal long-range order in this phase, represented by
\begin{equation}
\label{entangle}
\lim_{|{{\mathbf r}_\perp}'-{\mathbf r}_\perp|\rightarrow \infty}
\langle\psi({\mathbf r}_\perp,z)\psi^*({{\mathbf r}_\perp}',z')\rangle \neq 0
\end{equation}
where $\psi$ and $\psi^*$ are head and tail ``destruction'' and ``creation'' 
operators~\cite{Nelson:directed}, implying
entanglement of lines on a macroscopic scale. If defects are absent or appear 
only in closed loops,
the expression above would vanish as $|{{\mathbf r}_\perp}' - 
{\mathbf r}_\perp| \rightarrow \infty$. Once defects proliferate, 
a line can wander to any other column and Eq.~(\ref{entangle}) has a finite 
limit. A crystal with proliferating vacancies and interstitials is an 
incommensurate
phase --- the magnitude of the smallest reciprocal vector
$G = 4\pi/\sqrt{3}a_0$
is no longer related to the areal density in the obvious way as
$\rho = \sqrt{3}G^2/8\pi^2$ because the 
density differs from its defect-free value $\rho_0 = 2/\sqrt{3}a_0^2$ ($a_0$ 
being the lattice constant of the triangular lattice in cross-section). All 
crystals of pointlike atoms or molecules are trivially ``incommensurate'' in 
this sense --- the corresponding pointlike vacancies and interstitials 
proliferate at any finite temperature. It is the anomalous suppression of 
vacancies and interstitials and their organization into lines at low 
temperatures in columnar crystals which makes these materials unusual.

In this paper we apply the physics of directed lines to vacancy/interstitial 
strings. With this in mind, we briefly review the elasticity theory of these 
systems in the next Section. In Sec. \ref{single} we model a single string and 
estimate its transverse wandering. The form of this wandering is 
unchanged by coupling to phonon distortions of the lattice, as shown in 
Appendix \ref{bend}. So is its magnitude, as calculated in Appendix \ref{DR}. In
Sec. \ref{many} we apply the statistical mechanics of living polymers to an 
ensemble of directed strings and calculate their volume fraction, average 
length, 
etc. in the non-interacting limit. A simple quadratic-interaction model is 
presented in Section \ref{int}, similar to the one discussed via the boson 
mapping in 
Ref.~\cite{Nelson:directed}, and we reproduce the results therein. Numerical 
calculations of the line tensions of various species of defects are presented 
in Sec. \ref{num}. The interaction potentials considered are repulsive and 
monotonic; we study simple power laws as well as a screened Debye-H\"{u}ckel 
interaction. We find many metastable species of
vacancies. However, the lowest energy defect is always found to be the one with 
the highest symmetry in its category. For very short range interactions, this 
is the symmetric vacancy ($V_6$), whereas for most interactions the centered 
interstitial ($I_3$) is most favoured. Appendix \ref{Ewald} contains details of 
the Ewald summation calculations for the potentials considered here.

\section{Review of elasticity theory}
\label{theory}

Before discussing defects in a columnar crystal, we review the aspects of 
elasticity theory common to all the systems mentioned in the Introduction. We 
consider lines or columns aligned along a common direction ($\hat{\mathbf z}$) 
up to thermal fluctuations, with crystalline order in any cross-section 
perpendicular to the columnar axis.
In the case of flux lines, the average direction of alignment is imposed by an 
external field (${\mathbf H}=H \hat{\mathbf z}$) and local deviations from this 
direction cost energy.
With columnar crystals of long-chain molecules composed of covalently bonded 
nematogens or disk-shaped molecules cylindrically stacked via hydrogen bonds, 
or amphiphilic molecules in cylindrical micellar aggregates, the columnar axis 
represents spontaneously broken rotational symmetry.
Therefore local deviations from the alignment direction are not penalized, but 
undulations of the column are. The rotational symmetry can, however, be broken 
by imposing an external field. In addition, the two-dimensional crystalline 
order resists shear and areal deformations perpendicular to the
$\hat{\mathbf z}$-axis.

Low-energy fluctuations of the system can be described by a ``continuum'' model 
that works for small amplitude, long-wavelength 
deformations~\cite{SelB,Nelson:directed,deGennes}. 
The important fluctuations in this limit can be characterized by a 
two-dimensional displacement field ${\mathbf u}({\mathbf r}_\perp,z)$, 
representing the average deviation of lines in the ($x,y$) plane in a small 
region centered at $({\mathbf r}_\perp,z)$. With it can be associated a local 
areal density change
$\delta \rho/\rho_0 = - {\mathbf \nabla}_\perp \cdot {\mathbf u}$
($\rho_0 = 2/\sqrt{3} a_0^2$) and a local nematic director
${\hat{\mathbf n}} = {\hat{\mathbf z}} + {\mathbf t}$,
with ${\mathbf t} \equiv \partial{\mathbf u}/\partial z$.
The free energy of the system is a sum of nematic and crystalline contributions:
\begin{equation}
\label{Ftot}
{\mathcal F} = {\mathcal F}_{nematic} + {\mathcal F}_{crystal},
\end{equation}
To the lowest order in the fluctuations, these are given by
\begin{equation}
{\mathcal F}_{nematic} = \frac{1}{2} \int\!\! d^{3}r
		\left[K_1 ({\mathbf \nabla}_\perp \cdot {\mathbf t})^2
		 + K_2 ({\mathbf \nabla}_\perp \times {\mathbf t})^2 
 			+ K_3 (\partial_z {\mathbf t})^2 \right]
\end{equation}
and
\begin{equation}
\label{Fxtal}
{\mathcal F}_{crystal} =
	\int\!\! dz\! \int\!\! d^2{\mathbf r}_\perp 
	\left[\mu\,{u_{i j}}^2 +
	\frac{1}{2} \lambda\,\left(\frac{\delta\rho}{\rho_0}\right)^2 \right] 
\end{equation}
where $K_1, K_2, K_3$ are the Frank constants for splay, twist and bend 
respectively, and $\lambda$ and $\mu$ are the Lam\'{e} coefficients. The matrix 
$u_{ij} = (\partial_i u_j + \partial_j u_i)/2$ is the linearized 2D strain 
field.
In the presence of an external field $H {\hat{\mathbf z}}$, one should add to 
${\mathcal F}$:
\begin{equation}
\label{ext}
{\mathcal F}_{ext} = \frac{1}{2} \chi_a H^2 \int\!\! dz
\int\!\! d^2r_\perp |{\mathbf t}|^2 ,
\end{equation}
where $\chi_a$ is the anisotropic part of the susceptibility~\cite{deGennes}.

The last two contributions to ${\mathcal F}$ are quadratic in the derivatives, 
and can be rewritten as
\begin{eqnarray}
{\mathcal F}_{crystal} +
{\mathcal F}_{ext} = & \frac{1}{2} \int\!\! d^{3}r
	\left[c_{11} ({\mathbf \nabla}_\perp \cdot {\mathbf u})^2
		+ c_{66} ({\mathbf \nabla}_\perp \times {\mathbf u})^2
		+ c_{44} (\partial_z{\mathbf u})^2 \right]	\nonumber\\
		& \mbox{	} + \mu \;(\mbox{surface terms})
\end{eqnarray}
where $c_{11} \equiv \lambda + 2 \mu$, $c_{66} \equiv \mu$, and
$c_{44} \equiv \chi_a H^2 \rho$. The surface terms become important when there 
are defects within the bulk of the crystal, like vacancy/interstitial strings, 
represented by cuts joining column-end singularities in the field
${\mathbf u}({\mathbf r}_\perp,z)$. Evaluating these terms over a cylindrical 
surface enclosing such a string yields the energy cost of the defect string: a 
line tension $\tau_z \approx \mu a^2$ due to the elastic distortion around the 
string, in addition to a core energy $E_c$ per unit length (of the same order 
of magnitude) within the cylindrical core.

${\mathcal F}_{nematic}$ can be further simplified if, as is often the case 
with nematic polymers, the splay and twist constants are small in comparison to 
the bend constant. Specifically, if $K_1$ and $K_2$ satisfy
$K_{1,2} a_0^{-1}/\sqrt{K_3 c_{11}} \ll 1$~\cite{Jain:ice}, then they can be 
neglected. For long-wavelength distortions along the columnar axis, the 
dominant free energy contribution is then $K_3 (\partial_z^2 {\mathbf u})^2$ in 
the absence of an external field. $K_3$ can be simply related to the 
persistence length $l_P$ of the polymer as $K_3 = k_B T l_P \rho$.

\noindent
\begin{minipage}{2in}
\begin{figure}
\centering
\leavevmode
\epsfxsize=2in
\epsfbox{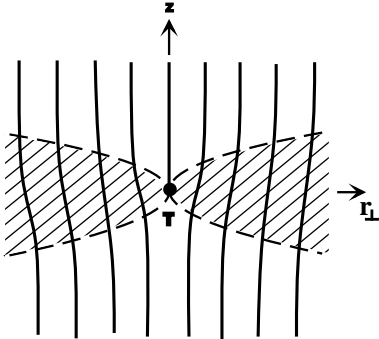}
\caption{Distortion induced by a column end in the neighbouring columnar 
crystalline matrix. The distortion is confined to a vertical extent
$|z| < \sqrt{\lambda_L r_\perp}$ (shaded region) around the column end.}
\label{distort}
\end{figure}
\end{minipage}
\hfill
\newlength{\textw}
\setlength{\textw}{\textwidth}
\addtolength{\textw}{-2.3in}
\begin{minipage}{\textw}
The statistical mechanics of defects in polymer liquid crystals has been 
discussd in detail by Selinger \& Bruinsma~\cite{SelB:defect,Meyer:defect}.
The presence of defects imposes a deformation on the $T=0$ equilibrium 
configuration. In the case of a semi-infinite vacancy/interstitial string with 
a head or tail at the origin, this distortion follows from minimization of the 
free energy above with respect to ${\mathbf u}({\mathbf r}_\perp,z)$ under the 
constraint
\begin{equation}
\label{constraint}
{\mathbf \nabla}_\perp \cdot {\mathbf u} = \pm \rho_0^{-1}
\delta({\mathbf r}_\perp) \theta(z) + (\mbox{non-singular terms})
\end{equation}
where the $\pm$ sign refers to a column tail/head located at the origin. Since 
the planar distortion about a string has azimuthal symmetry in the continuum 
approximation, ${\mathbf \nabla}_\perp \times {\mathbf u} = 0$. Hence, the only 
relevant terms in the free energy are the bend and bulk distortion terms 
(neglecting splay). The resulting distortion around the column end spans a 
parabolic region about the radial direction (see Fig.~\ref{distort}) defined by
\begin{equation}
\label{lambda}
z^2 \lesssim \lambda_L r_\perp
\end{equation}
where $\lambda_L = \sqrt{K_3/c_{11}}$
is the length scale relating the distortions parallel and perpendicular to 
${\hat{\mathbf z}}$.
\end{minipage}
\medskip

Selinger \& Bruinsma also calculate the interaction energy between two column 
ends by superimposing the distortion created by each. They find the interesting 
result that a head and tail in a \textit{nematic} medium attract weakly if they 
fall within each other's region of influence, as just described, but repel 
otherwise. However, in a columnar crystal (with non-zero shear modulus), the 
interaction is always a strong attractive linear potential due to the finite 
line tension associated with the string of distortions joining a head to a tail.


\section{Wandering of a single string}
\label{single}
Consider a single vacancy/interstitial string in a hexagonal columnar crystal 
of, say, polymer strands with lattice constant $a_0$ and monomer spacing $c$ 
along the columnar axis 
$\hat{\mathbf z}$. For a discotic columnar liquid crystal, $c$ is the spacing 
between oblate molecules along the column axis. For a flux line in a layered 
Type-II superconductor with magnetic field perpendicular to the layers, $c$ is 
the layer spacing. If the string is vertical, the energy per unit length 
$\tau_z$ is of the order of $\mu a_0^2$ (see Section \ref{theory}) where $\mu$ 
is the in-plane shear modulus 
of the crystal.  For a horizontal string, $\tau_\perp = \varepsilon_k/a_0$ where
the kink energy
$\varepsilon_k \sim \kappa^{1/4} \mu^{3/4} a_0^2$~\cite{Nelson:directed},
$\kappa \equiv K_3/\rho$ being the bending rigidity. The ratio is
$\tau_\perp/\tau_z \sim (\kappa/\mu)^{1/4}/a_0 \sim l^*/a$ where $l^*$ is the 
kink size. Typically $l^* \gg a_0$, so that the strings are predominantly 
vertical, with few kinks. For flux lines on the other hand, the kink energy is 
$g^{1/2} \mu^{1/2} a_0$ with $g \equiv c_{44}/\rho$, where $c_{44}$ is the tilt 
modulus and $\rho$ is the areal line density. The ratio is then 
$(g/\mu)^{1/2}/a_0$. In highly anisotropic layered superconductors, this ratio 
can be small, favouring large, nearly horizontal defect excursions. We will for 
now work with nearly vertical strings, allowing for a gas of kinks sufficiently 
dilute so that the interaction between kinks can be ignored (see 
Fig.~\ref{walk}). We thus assign to a string of vertical extent $l$ and $n_k$ 
kinks an energy $l \tau + n_k \varepsilon_k + 2 \varepsilon_0$ where
$\tau \equiv \tau_z$ and $\varepsilon_0$ is the energy of a polymer end. We 
expect that the results for defects with a high density of kinks would be 
qualitatively similar.

In units such that $k_B=1$, the partition function of a string of length $l$ is
\begin{equation}
{\mathcal Z}_1 = (1 + q e^{-\varepsilon_k/T})^{l/l^*} e^{-l \tau/T}
\end{equation}
where $T$ is the temperature, and $q$ is the 
two-dimensional co-ordination number of the lattice on which the defect string 
lives --- for a symmetric vacancy this is the same as that of the original 
triangular lattice, $q = 6$, 
whereas for a symmetric interstitial it is that of the dual honeycomb lattice, 
$q = 3$ (see Section \ref{num}). The above 
expression represents the freedom of the string to jump to any of the 
neighbouring lattice sites anywhere along its length. These transverse 
meanderings cause an entropic lowering of the free energy per unit length of 
the string:
\begin{eqnarray}
f_1	&=& \lim_{l \rightarrow \infty} -T \ln{{\mathcal Z}_1} /l	\nonumber	\\
	&=& \tau - \frac{T}{l^*} \ln{\left(1 + q e^{-\varepsilon_k/T}\right)}
	\nonumber \\
	&\simeq& \tau - \frac{T q}{l^*} e^{-\varepsilon_k/T} \quad\mbox{for}\quad
	e^{-\varepsilon_k/T} \ll 1
\end{eqnarray}
If $N_k$ is the total number of kinks, the average kink density is
\begin{eqnarray}
n_k \equiv \frac{\langle{N_k}\rangle}{l} &=& \frac{1}{l^*}
\frac{q e^{\varepsilon_k/T}}{1+q e^{\varepsilon_k/T}} \nonumber \\
	&\simeq& \frac{q}{l^*} e^{-\varepsilon_k/T}, \quad\mbox{for}\quad
	e^{-\varepsilon_k/T} \ll 1.
\end{eqnarray}
Thus, kinks are on the average $l_k = l^* e^{\varepsilon_k/T}/q$ monomers apart.
The assumption of dilute kinks then translates into the condition $l^* n_k \ll 
1$, or, $\varepsilon_k \gg T$, which can be rephrased as
$\langle |{\mathbf u}|^2 \rangle/a_0^2 \ll 1$~\cite{Nelson:directed,Jain:ice}, 
a condition clearly satisfied by a crystal below its Lindemann melting point.

The above is a ``diffusive'' model for the string --- if ${\mathbf d}$ denotes 
the 
horizontal end-to-end displacement, the mean square wandering is 
$\langle|{\mathbf d}|^2\rangle = 2 D l$, where the ``diffusion constant'' $D$ 
is given by $2 D = a_0^2 n_k$. Consider a continuum description of the string 
in terms of a function ${\mathbf r}_d(z)$, ${\mathbf r}_d(z)$ being the 
transverse displacement. Provided the average slope $|d{\mathbf r}_d/dz|$ is 
small, this ``diffusive'' wandering would correspond to an effective 
Hamiltonian of the form
\begin{equation}
\label{defEg}
H_1 = \int_0^l dz \left[\frac{g}{2} \left|\frac{d{\mathbf r}_d}{dz}\right|^2 + 
\tau\right], \quad g=\frac{T}{D}
\end{equation}

Here we have assumed that 
the string is wandering within a frozen crystal. However, the lattice around 
the vacancy/interstitial string responds to its presence by collapsing or 
expanding
around it. For a straight string at ${\mathbf r}_d = {\mathbf 0}$, the deformation 
${\mathbf u}({\mathbf r}_\perp,z)$ is given by
\begin{equation}
{\mathbf u}_d({\mathbf r}_\perp,z) =
\pm \frac{\Omega}{2 \pi} \frac{{\mathbf r}_\perp}{r_\perp^2}
\end{equation}
in the continuum description of the crystal, that is, away from the defect 
where the deformations are small. $\Omega$ is the area change due to the 
vacancy/interstitial, $\Omega \simeq a_0^2$. The energy of this 
deformation has to be included in the energy cost of the defect string. 
Again invoking the continuum approximation, we assume that for a defect string 
with small average slope, the resulting deformation away from the string in any 
plane perpendicular to $\hat{\mathbf z}$ would be approximately that resulting 
from a straight string at the 
location of the defect in that plane:
\begin{equation}
{\mathbf u}({\mathbf r}_\perp,z) \simeq
{\mathbf u}_d({\mathbf r}_\perp-{\mathbf r}_d(z),z).
\end{equation}
(In general ${\mathbf u}({\mathbf r}_\perp,z)$ would depend on the derivatives 
of ${\mathbf r}_d(z)$ as well.) Within this approximation, the distortion energy
of the crystal with bending Frank's constant 
$K_3 \equiv T l_P \rho$ is, keeping terms up to fourth-order in the derivatives 
(see Appendix \ref{bend}):
\begin{equation}
\label{defE4}
\frac{\Delta H_1}{T} \sim
l_P \int dz \left[ \left|\frac{d^2 {\mathbf r}_d}{dz^2}\right|^2,
a_0^{-2} \left|\frac{d{\mathbf r}_d}{dz}\right|^4 \right]
\end{equation}
These impart an effective stiffness to the defect string and suppress 
transverse fluctuations over a length scale $\sim a_0\sqrt{D K_3/T} \sim 
a_0\sqrt{l_P n_k}$. However, they do not change the long 
scale diffusive nature of the string.

The lattice distortions renormalize the diffusion constant of the string when 
the
symmetry direction of the crystal is externally imposed, as in the case of 
flux lines, or in a polymer crystal with an external field along the 
$\hat{\mathbf z}$-direction. The tilt modulus $c_{44}$ is then non-zero 
(Eq.~(\ref{ext})), and D is renormalized to $D_R$, where (see Appendix \ref{DR})
\begin{equation}
\frac{1}{D_R} \simeq \frac{1}{D} +
{\mathcal O}\left(\frac{c_{44}}{T \rho}\right)
\end{equation}

For a dense vortex 
\textit{liquid} this effect has been analyzed in detail by 
Marchetti~\cite{Marchetti:D} 
and $D$ is found to be renormalized to a value independent of its bare value in 
the long-wavelength limit. The correction comes from convection of a tagged 
flux line along the local tangent-field direction.

If a similar calculation is carried out for a \textit{crystal} of spontaneously 
aligned 
long semi-flexible polymers (see Appendix \ref{DR}),
one finds a qualitatively different renormalization of $D$ --- the correction 
in the long-wavelength limit is proportional to its bare value,
and $\delta D/D \sim 1.45 \langle|{\mathbf u}|^2\rangle/a_0^2 \lesssim 3~\%$
using $c_L^2 \simeq 1/50$~\cite{cl} ($c_L$ is the Lindemann constant for 
melting of a 
columnar crystal). The correction is negligible. It can be ignored for 
another reason --- the idea of convection of a line by the mean local field, 
although appropriate for a dense fluid, would not be applicable in a 
crystalline environment where diffusion can only occur through discrete jumps 
from column to column.
Although thermal fluctuations are already implicit in the exponential factor 
in $D = a_0^2 n_k/2$ coming from $n_k$, defects in this case move only on 
a discrete lattice, without phonon fluctuations.

To summarize this section, we characterize the statistical mechanics of a 
defect string with a head/tail energy 
$\varepsilon_0$, a line tension $\tau$, and a diffusion constant $D$. The 
latter two can be combined in an effective chemical potential $\overline{\mu} 
\equiv T \mu_d$ per kink size ($l^*$) of the string:
\begin{equation}
\mu_d = l^* (-\tau/T + n_k) = q e^{-\varepsilon_k/T} - \varepsilon_k/T,
\end{equation}
with $n_k$ related to $D$ through $D = a_0^2 n_k/2$. Because $n_k$ is 
exponentially 
small, $\mu_d \approx -l^* \tau/T \approx -l^* \mu a_0^2/T$ and is usually 
negative, which suppresses long vacancy \& interstitial strings. Turning it 
positive would require raising 
the temperature and lowering the kink energy $\varepsilon_k$, and is favored by 
a larger co-ordination number $q$.

Although we have assumed a constant shear modulus, the presence of the defects 
themselves can drive it down exponentially with the defect concentration, as 
discussed by Carruzzo \& Yu~\cite{CarYu:shear}. Thus, positive $\mu_d$ becomes 
possible when softening of the bare elastic constants with increasing defect 
concentration is taken into account.

\section{Statistical Mechanics of non-interacting strings}
\label{many}
At any finite temperature, a crystal with a negative string line-chemical 
potential will contain a distribution of thermally excited vacancy and 
interstitial strings. Since the string energy is proportional to length in the 
non-interacting-kinks approximation, the equilibrium probability distribution 
would be an exponentially decaying function of length with mean determined by 
the line chemical potential, in the dilute string-gas limit where
inter-string interactions can also be neglected~\cite{Safran}. In discotic 
crystals string heads and tails can be created as necessary. In a crystal of 
long polymers, the number of heads and tails is fixed by the mean polymer 
length.

Let N be the total number of possible kink sites in the lattice, 
$N = \mbox{volume} \times \rho/l^*$, and ${\mathcal P}_l$ be $1/N \times$ the 
number of defect strings $l$-links long. Assuming that only one kind of defect
string is present --- those with the lowest line tension --- we can write the 
defect 
free energy in terms of $\{{\mathcal P}_l\}$ as~\cite{Safran}
\begin{equation}
\label{F}
{\mathcal F}_d(\{{\mathcal P}_l\}) =
\sum_l N {\mathcal P}_l (2 \varepsilon_0 - l T \mu_d) +
T \sum_l N {\mathcal P}_l (\ln{{\mathcal P}_l} - 1)
\end{equation}
Minimizing with respect to the $\{{\mathcal P}_l\}$ yields the expected 
exponential distribution:
\begin{equation}
{\mathcal P}_l = h^2 z^l
\end{equation}
where $z = e^{\mu_d}$, and the head/tail fugacity $h = e^{-\varepsilon_0/T}$ is 
expected to be small. For hexagonal columnar crystals of \textit{polymers}, we 
work in a grand canonical ensemble and adjust $\varepsilon_0$ so that the 
average head/tail concentration agrees with the fixed value determined by the 
mean polymer length. The head/tail concentration will be small if the polymers 
are long. For \textit{discotic} crystals, the grand canonical ensemble is the 
natural one and the head/tail concentration fluctuates, with an average value 
determined by the fixed value of $h = e^{-\varepsilon_0/T}$,
and the monomer fugacity $z = e^{\mu_d} < 1$.
The net defect volume fraction $\phi$ is
\begin{equation}
\phi \equiv \sum_l l {\mathcal P}_l = h^2 \frac{z}{(1-z)^2} .
\end{equation}
The total number of strings $N_d \equiv N n_s$ is given by the string density
\begin{equation}
n_s \equiv \sum_l {\mathcal P}_l = \frac{h^2}{1-z}
\end{equation}
A defect monomer is most likely to be found in a string of mean length (in 
units of the kink size)
\begin{equation}
l_m = \frac{1}{|\mu_d|}
\end{equation}
The length distribution has an average at $2 l_m$, and a spread also of 
$\sqrt{2} l_m$.
The form (\ref{F}) of the energy, linear in $l$, is really applicable only when 
$l \gg 1$, so that end effects can be parametrized by the $l$-independent 
constant $\varepsilon_0$. Then, $\mu_d$ is close to $0$, and the relation
$\phi \simeq n_s l_{mp}$ holds.
The asymptotic behaviours in the dilute and dense limits are as follows:
\begin{eqnarray}
\phi &=& \left\{\begin{array}{ll}
	h^2 e^{\mu_d}, & z \ll 1 \\
	\frac{h^2}{|\mu_d|^2}, & z \lesssim 1
	\end{array}\right. \\
n_s &=& \left\{\begin{array}{ll}
	h^2, & z \ll 1 \\
	\frac{h^2}{|\mu_d|}, & z \lesssim 1
	\end{array}\right.
\end{eqnarray}

A string proliferation transition thus occurs at $\mu_d = 0$ in this model, 
corresponding to a temperature $T_d = \tau l_k$. In the limit
$\varepsilon_0 \rightarrow \infty$, it corresponds to the appearance of a 
supersolid
phase~\cite{Frey:defect} which is simultaneously crystalline and entangled, 
where infinitely long 
vacancy/interstitial strings facilitate the wandering and entanglement of lines 
in the crystalline phase. If the melting temperature $T_m > T_d$, this 
supersolid/incommensurate solid phase will exist between $T_d$ and $T_m$.

The non-interacting approximation breaks down in the vicinity of $T_d$ as 
calculated here, and its estimate will have to be refined by including 
interactions. For finite $\varepsilon_0$, the sharp transition discussed in Ref.
~\cite{Frey:defect} will be blurred, as discussed in Sec. \ref{int}.

\section{$\phi^2$-interaction model}
\label{int}
Interactions between polymer ends in a columnar crystal have been calculated by 
Selinger \& Bruinsma~\cite{SelB:defect} within the continuum approximation. 
Because 
of the uniaxial anisotropy, the interaction has a rather complicated 
form. The distortion due to an isolated head or tail placed at the origin at 
in-plane distance $r_\perp$ extends 
over a vertical extent $|z| \sim \sqrt{\lambda_L r_\perp}$ where
$\lambda_L = \sqrt{K_3/c_{11}}$ (see Eq.~(\ref{lambda})). The resulting 
interaction between heads and tails falls as $1/|z|^3$ 
for predominantly vertical separations $z$ ($|z| \gg \sqrt{\lambda_L r_\perp}$),
and as $-1/(\lambda_L r_\perp)^{3/2}$ for predominantly horizontal separations 
$r_\perp$. In polymer crystals, these contributions must be superimposed on the 
linear energy cost of the vacancy or interstitial string joining them.

At low defect densities where the string length is much smaller than the average
separation of string centers of mass, we have $1/|\mu_d| \ll 1/\phi^{1/3}$, 
i.e., $|\mu_d| \gg h^{2/3}$, and
a string interacts with other strings as a head-tail dipole. The effective 
 interaction between dipoles then falls off very rapidly, becoming short-ranged 
 not only in the axial, but also in the radial direction.

At the other extreme, the strings are long, which would happen in the vicinity 
of the head-tail unbinding transition and in the supersolid phase itself. 
End-interactions can then be neglected and the remaining interaction between 
effectively infinite strings becomes predominantly ``radial'' (i.e., 
perpendicular to $\hat{\mathbf z}$) provided the root mean square tilt with 
respect to the $\hat{\mathbf z}$ axis is small.
The defects are 
then non-interacting in the continuum model unless their anisotropy is taken
into account. The interaction between defects with n-fold symmetry (n = 2, 3 or 
6) falls off at least as fast as $1/r^n$ (see Appendix \ref{n-def}).
This interaction has an azimuthal dependence of the 
form $\cos{n \theta}$ or higher harmonics. The angular average vanishes,
leading to an effective interaction which vanishes as an even higher power 
which is effectively short-ranged. As mentioned in the Introduction, the 
lowest-energy vacancy or interstitial defects for simple repulsive pair 
potentials in the radial direction are in fact of high (three-fold or six-fold) 
symmetry.

We discuss here the simplest model for a short-ranged interaction --- a 
repulsive $\phi^2$ model that has been treated earlier in 
Ref.~\cite{Nelson:directed} using a 
coherent state path integral representation which exploits an analogy with the 
quantum mechanics of two-dimensional bosons. The defect volume fraction $\phi$ 
corresponds to 
the mean square boson field amplitude $\langle|\psi|^2\rangle$ in that 
description. Here, we reproduce the essential results without resorting to the 
sophisticated boson formalism.
Upon adding a term $u \phi^2 /2$ to the free energy $f \equiv F/N T$ in 
Eq.~(\ref{F})
of the previous section, we find after minimization, 
\begin{equation}
{\mathcal P}_l = h^2 e^{l (\mu_d - u \phi)} .
\end{equation}
As discussed in Ref.~\cite{Nelson:directed}, the coupling $u$ is an excluded 
volume parameter describing defect line repulsion. Thus $\phi$ and $N_d$ have 
the same form as before, but with $z$ replaced by an effective fugacity $\zeta$:
\begin{equation}
\label{zeta}
z \rightarrow \zeta(z,\phi) \equiv z e^{-u \phi}  ,
\end{equation}
so that
\begin{equation}
\label{phi-zeta}
\phi(h,\zeta) = h^2 \frac{\zeta}{(1-\zeta)^2} .
\end{equation}
The volume fraction $\phi(h,z)$ now has to be solved for self-consistently from 
Eq.~(\ref{phi-zeta}). Note that the effective chemical potential has been 
reduced by 
$u \phi$ due to the repulsive interaction:
\begin{equation}
\mu_{eff} \equiv \ln{\zeta} = \mu_d - u \phi
\end{equation}
Accordingly, the mean string length $l_m$ changes to 
\begin{equation}
 l_m = -\frac{1}{\ln{\zeta}} \equiv \frac{1}{u \phi - \mu_d}  .
\end{equation}
The free energy of the distribution is $f \approx -u \phi^2 /2$.

The behaviour of the string volume fraction for $h = 0$ and $h \neq 0$ is 
illustrated schematically in Fig.~\ref{phimu}. Four distinct regimes emerge, 
with the following asymptotic behaviours:

\begin{figure}
\centering
\leavevmode
\epsfxsize=6.1in
\epsfbox{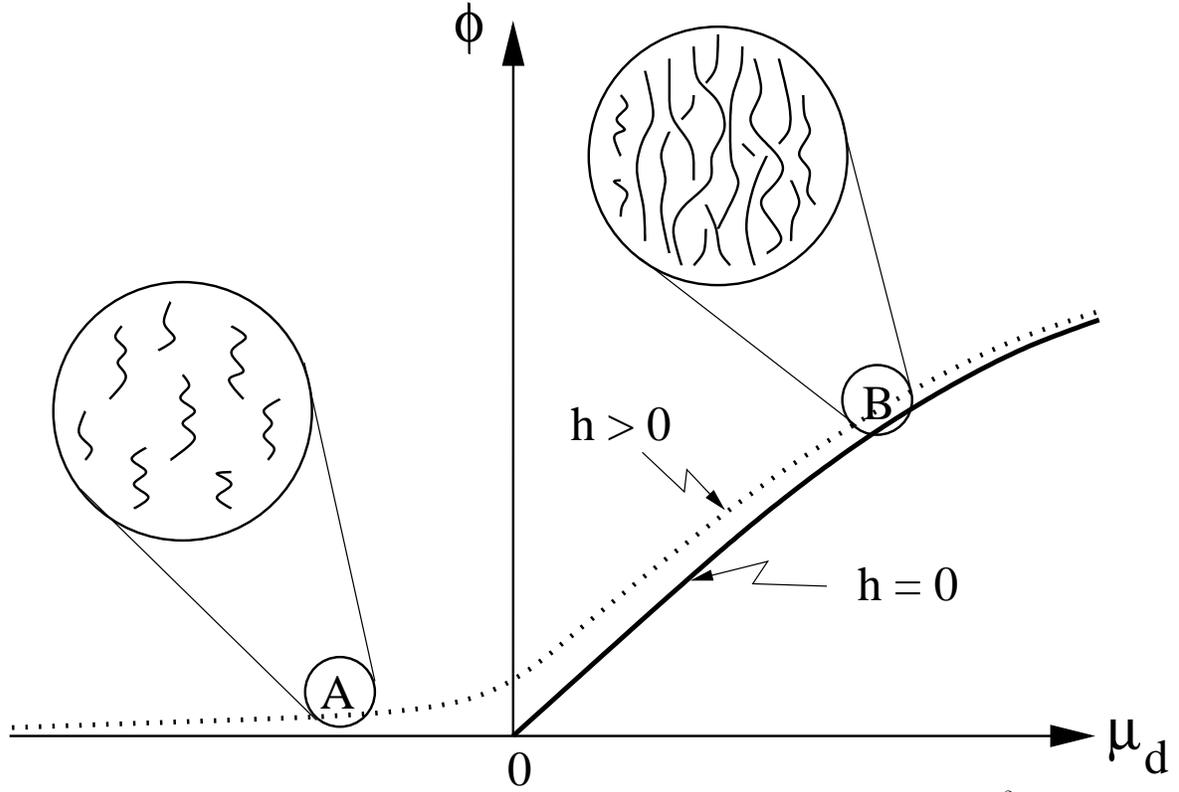}
\caption{The volume fraction $\phi$ is plotted against the effective defect 
chemical potential $\mu_d$ for the $\phi^2$-interaction model of a gas of 
defect strings. The strings are short and dilute in regime A, but long, dense 
and entangled in regime B. (Taken from Ref.~\protect\cite{Nelson:directed}.)}
\label{phimu}
\end{figure}

\begin{enumerate}

\item $\mu_d \ll -1$ (point A in Fig.~\ref{phimu}):
\begin{equation}
\phi \simeq h^2 e^{\mu_d},\quad
n_s \simeq h^2,\quad
l_m = \frac{1}{|\mu_d|} .
\end{equation}
This is again the dilute limit where heads and tails are tightly bound.

\item $-1 \ll \mu_d \ll -(u h^2)^{1/3}$:
\begin{equation}
\label{noni}
\phi \simeq \frac{h^2}{|\mu_d|^2},\quad
n_s \simeq \frac{h^2}{|\mu_d|},\quad
l_m = \frac{1}{|\mu_d|} .
\end{equation} 
These results are again identical to those for non-interacting strings. This 
correspondence is expected, because $|\mu_d| > (u h^2)^{1/3} > u \phi$, 
therefore the effective chemical potential is still approximately $\mu_d$. The 
relation $\mu_d \sim -(u h^2)^{1/3}$ marks the limit of validity of the 
non-interacting approximation, as we argued in the beginning of this section. 
As we approach this limit, we find for $h \rightarrow 0$:
$\phi, n_s \rightarrow 0$, whereas $l_m \rightarrow \infty$. Thus, the strings 
are still dilute, although lengthening. 
Note that the results in this regime coincide with those of 
Ref.~\cite{Nelson:directed} in the short and dilute strings limit.

\item $|\mu_d| \ll (u h^2)^{1/3} \equiv \mu_c$ ($\mu_d$ around the transition 
which occurs for $h = 0$):
\begin{equation}
\phi \simeq
\frac{h^2}{|\mu_c|^2} \left[ 1 + \frac{2}{3}\frac{\mu_d}{\mu_c}\right],\quad
n_s \simeq
\frac{h^2}{|\mu_c|} \left[ 1 + \frac{1}{3}\frac{\mu_d}{\mu_c}\right],\quad
l_m \simeq
\frac{1}{|\mu_c|} \left[ 1 + \frac{1}{3}\frac{\mu_d}{\mu_c}\right] .
\end{equation}
These results can be matched onto those in the non-interacting regime above by 
replacing $\mu_d$ with
\begin{equation}
\mu_{eff} = -\mu_c + \mu_d/3 = -\mu_c \left(1-\frac{\mu_d}{3 \mu_c}\right) ,
\end{equation}
which is now dominated by the repulsive interaction:
$\mu_{eff} \approx - u \phi$. The unphysical divergences of the non-interacting 
model have been suppressed and we find at the transition point:
\begin{equation}
\phi = \frac{h^{2/3}}{u^{4/3}},\quad
n_s = \frac{h^{4/3}}{u^{1/3}},\quad
l_m = \frac{1}{u^{1/3} h^{2/3}}  .
\end{equation}
Note that all quantities have interesting singularities in the limit
$h \rightarrow 0$.

If the head/tail fugacity $h$ is small, the defect volume fraction remains 
negligible at the transition, but the average string length grows large so that 
it could become greater than the inter-string separation, now given by 
$1/\phi^{1/2}$.
Indeed, $1/\phi^{1/2} \ll l_m$ if $h \ll 1/u^2$ which would be true if polymer 
ends are highly unfavourable.

This long \& dilute regime interpolates between the short \& dilute and the 
long \& dense limits described in Ref.~\cite{Nelson:directed}.

\item $\mu_d \gg \mu_c$ (Point B in Fig.~\ref{phimu}):\\
In this limit, we have
\begin{equation}
\mu_{eff} = -\mu_c \sqrt{\frac{\mu_c}{\mu_d}} .
\end{equation}
The repulsion now keeps in check the string proliferation, and $\mu_{eff}$ 
approaches $0$ as $1/\sqrt{\mu_d}$. Thus,
\begin{equation}
\phi \simeq \frac{\mu_d}{u},\quad
n_s \simeq h \sqrt{\frac{\mu_d}{u}},\quad
l_m \simeq \frac{1}{|\mu_c|} \sqrt{\frac{\mu_d}{\mu_c}}  .
\end{equation}
This is the phase where strings are dense and entangled --- $\phi$ is 
${\mathcal O}(1)$. These results also agree with Ref.~\cite{Nelson:directed}. 
\end{enumerate}

As the head/tail fugacity $h \rightarrow 0$, the intermediate regime~3 above 
(around $\mu = 0$) 
shrinks to zero. At $h = 0$, heads/tails are completely expelled, and we have a 
second-order phase transition at $\mu_d = 0$ with $\phi = 0$ for 
$\mu_d < 0$, and growing as $\mu_d$ for $\mu_d > 0$, as in 
Ref.~\cite{Nelson:directed}. This limit corresponds to the situation in 
thermally excited vortex lattices~\cite{Frey:defect} because flux lines cannot 
start or stop within the sample. In the boson picture, $h$ acts like an 
external field coupled to the order parameter, injecting magnetic monopoles 
into the superconductor.

We have neglected vacancy/interstitial loops, which exist even in the limit 
$h \rightarrow 0$. For finite $h$, their contribution can be neglected 
near the transition because for long loops, the energy of a loop exceeds 
the energy of a string of the same vertical extent: 
Whereas a string of length $l$ has energy
$l \tau_{interstitial} + 2 \varepsilon_0$ (we expect interstitials to be the 
preferred defect at the transition in most cases), the energy of a 
vacancy-interstitial loop of the same length would be approximately
$l\: (\tau_{vacancy} + \tau_{interstitial})$. For large $l$, the difference 
$l \tau_{vacancy} - 2 \varepsilon_0$
will strongly suppress vacancy/interstitial loops. Because of this energetic 
barrier, loops cannot become arbitrarily large, and cannot cause entanglement 
over macroscopic scales. For $h = 0$, as is the case for vortex matter, 
fluctuations in the low temperature phase are entirely in the form of 
loops~\cite{Frey:defect}, and similar to vortex ring fluctuations in the 
Meissner phase.

For systems with a finite axial length, the balance may be tilted in favour 
of long strings because the end penalty is removed if the ends 
move to the surface and the string threads the sample. For threading strings 
the expression for entropy in Eq.~(\ref{F}) is no longer valid because the 
freedom in the z-direction is lost. The remaining two-dimensional entropy can be
ignored in a three-dimensional system, and we are left with 
\begin{equation}
 f \simeq -\mu_d \phi + u \phi^2 /2 
\end{equation}
where $\phi$ now is also the areal fraction of defects; and one finds 
$\phi \simeq \mu_d/u$, similar to region~4 discussed above.

\section{Numerical calculation of defect line tensions}
\label{num}

Line tension calculations require that we find the lowest energy lattice 
deformation associated with a vacancy or interstitial. These line tensions 
depend on the \textit{type} of vacancy or interstitial, e.g., whether the 
defect sits in an environment which is two-, three- or six-fold symmetric. If 
thermal fluctuations out of this configuration are small enough to be described 
within a quadratic approximation, they decouple from the equilibrium 
configuration. Since these $T=0$ equilibrium defect configurations are composed 
of straight columns, the 3-dimensional deformation energy can be reduced to an 
effective 2-dimensional interaction energy $V(r)$ per unit length between 
columns separated by distance $r$. The calculations can then be performed on a 
two-dimensional triangular lattice of points interacting with potential $V(r)$. 
Thus, the defect energies in a two-dimensional Wigner crystal of 
electrons~\cite{Morf:defect} would correspond to the \textit{line tensions} of 
the corresponding string defects in a hexagonal columnar crystal of lines 
interacting with an effective radial $1/r$-potential per unit length.

\begin{figure}
\centering
\leavevmode
\epsfxsize=6.3in
\epsfbox{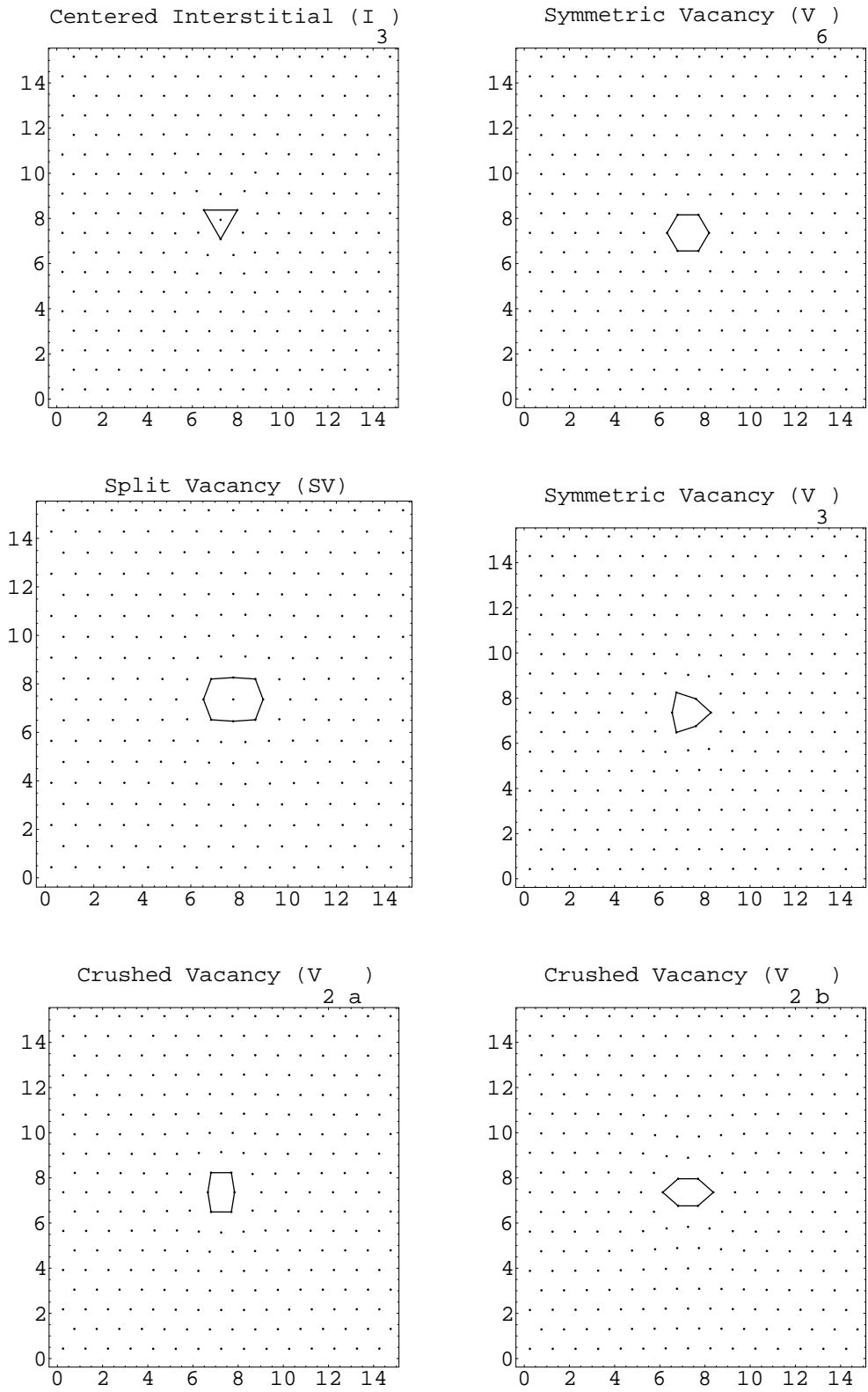}
\vfill
\caption{Various defects obtained in a two-dimensional triangular lattice. The 
centered interstitial is the only stable interstitial defect.}
\label{defects}
\end{figure}

Such calculations have been carried out by several 
authors~\cite{Frey:defect,Morf:defect,CockEl:defect}. Whereas 
Refs.~\cite{Morf:defect} and~\cite{CockEl:defect} have considered defects in a 
Wigner crystal of electrons ($V_p(r)=1/r$),
Frey \textit{et al.}~\cite{Frey:defect} have studied a modified Bessel-function 
potential $V_{\kappa}(r) = u_0 K_0(\kappa r)$ in the $\kappa \rightarrow 0$ 
limit. Here $\kappa \equiv \lambda^{-1}$, where $\lambda$ is the Debye 
screening length in the case of long polyelectrolytes in an ionic solution, and 
the London penetration depth in the case of vortex lines in a type-II 
superconductor. The limit $\kappa \rightarrow 0$ corresponds to a long-range 
logarithmic interaction, whereas in the short-range limit $\kappa a_0 \gg 1$ 
the interaction is exponentially decaying.
Both Refs.~\cite{Frey:defect} and~\cite{CockEl:defect} dealt with long-range 
interactions ($\ln{r}$ and $1/r$ repectively), and found that the centered 
interstitial (see Fig.~\ref{defects}) has the lowest line tension. We denote 
the centered interstitial by $CI$, or by $I_3$ when we want to stress its 
three-fold symmetry. The edge interstitial (denoted $EI$ or $I_2$) was found to 
be a saddle-point and buckled into a $CI$. The three-fold symmetric centered 
interstitial $CI$ is the lowest energy interstitial defect over the entire 
range of interactions we studied. Among the vacancies, the two-fold symmetric 
crushed vacancy (denoted $V_2$ or $V_{2a}$ --- see Fig.~\ref{defects}) is the 
only stable one, the symmetric six-fold vacancy ($V_6$) being unstable to it. 
The long-range interactions between the energetically preferred types of 
interstitials and vacancies were found to be attractive for interstitials and 
repulsive for vacancies.

To determine the correct type of microscopic defect to insert into the 
phenomenological considerations of Secs. \ref{single}--\ref{int}, we have 
extended the work of Frey \textit{et al.} to the short-ranged regime of the 
$K_0(\kappa r)$-interaction, to which end we studied values of $\kappa a_0$ 
from $0$ to $7$ ($7$ being large enough to represent the short-range
$\kappa a_0 \rightarrow \infty$ limit) (Fig.~\ref{bes:ek}). The aim was to 
determine the point of cross-over from centered interstitials to vacancies as 
the lowest-energy defect, since it is known from simulations of short-range 
interactions (for a review, see Ref.~\cite{point-defects}) that vacancies are 
preferred in this limit. In the same spirit, we have also extended the Coulomb 
interaction to power-law interactions $1/r^p$ with exponent values ranging from 
$p = 0$ ($\sim \ln{r}$) to $p = 12$ (Fig.~\ref{gam:ep}).

We checked our minimization procedure by first reproducing the results of 
Refs.~\cite{Frey:defect} and~\cite{CockEl:defect} for $\ln{r}$ and $1/r$ 
potentials respectively. As we move away from the long-range interaction limit 
$\kappa a = 0$, the metastable crushed vacancy ($V_{2a}$) exchanges stability 
with the metastable split vacancy ($SV$), also of two-fold symmetry. Two 
metastable species, a three-fold symmetric vacancy ($V_3$) and a two-fold 
symmetric vacancy ($V_{2b}$) crushed along the basis vector of a triangular 
unit cell, also exist, but are of higher energy. The differences in energy can 
be as small as one part in a few thousand. As the interaction gets 
shorter-ranged, $V_{2b}$ loses stability to $V_3$ at $\kappa a_0 \simeq 5.2$, 
and the 3-fold deformation of $V_3$ gets smaller so that it transforms 
continuously into $V_6$ at $\kappa a_0 \simeq 5.9$. When $V_6$ appears, the 
$SV$ also loses stability to it. By the time $I_3$ and $V_6$ finally cross in 
energy, $V_6$ is the only stable vacancy left. The crossing happens at 
surprisingly large parameter values, $\kappa a_0 \simeq 6.9$ for $V_{\kappa a}$ 
(Fig.~\ref{bes:diff}), and $p \simeq 5.9$ for $V_p$ (Fig.~\ref{gam:diff}), each 
very close to the short-range limit. We thus find that the interstitial has a 
very wide range of stability, extending well into the short-ranged regime.

\noindent
\begin{minipage}{3.4in}
\begin{figure}
\centering
\leavevmode
\epsfxsize=3.2in
\epsfbox{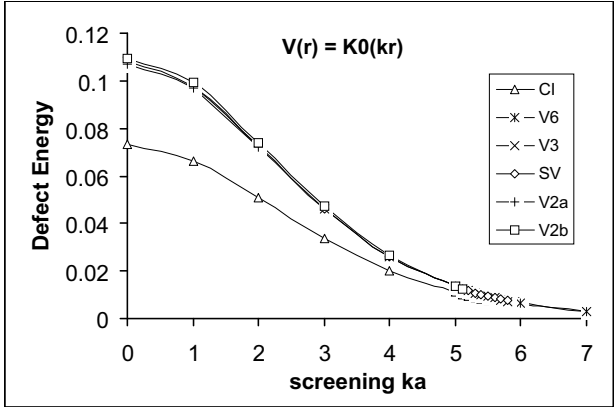}
\caption{Defect energy as a function of the screening $\kappa a$ for
$V(r) = K_0(\kappa r)$ at system size $n = 4$ ($N = 480$). Only the centered 
interstitial is shown, because the edge interstitial is always unstable to it. 
Various species of vacancies exist, within limited parameter ranges, very close 
in energy. Lines joining the data points are only an aid to the eye.}
\label{bes:ek}
\end{figure}
\end{minipage}
\hfill
\begin{minipage}{3.4in}
\begin{figure}
\centering
\leavevmode
\epsfxsize=3.2in
\epsfbox{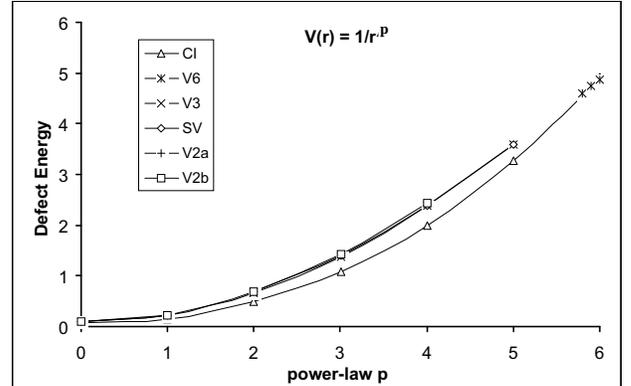}
\caption{Defect energy as a function of the screening $\kappa a$ for
$V(r) = 1/r^p$ at system size $n = 5$ ($N = 750$). The apparent increase in 
energy with $p$ (interaction getting shorter-ranged) would go away with proper 
normalization of the potential. Lines joining the data points are only an aid 
to the eye.}
\label{gam:ep}
\end{figure}
\end{minipage}
\medskip
\noindent
\begin{minipage}{3.4in}
\begin{figure}
\centering
\leavevmode
\epsfxsize=3.2in
\epsfbox{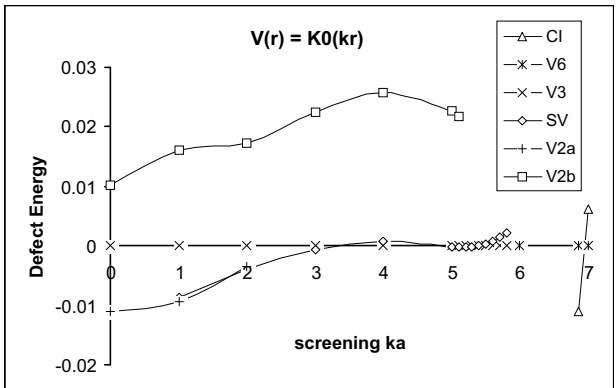}
\caption{Defect energies for $V(r) = K_0(\kappa r)$, $n = 4$, on the log-scale, 
with respect to $V_3$/$V_6$, in order to illustrate the detailed structure of 
the energy diagram. The $CI$ can be seen crossing $V_6$ at
$\kappa a \approx 6.9$. Lines joining the data points are only an aid to the 
eye.}
\label{bes:diff}
\end{figure}
\end{minipage}
\hfill
\begin{minipage}{3.4in}
\begin{figure}
\centering
\leavevmode
\epsfxsize=3.2in
\epsfbox{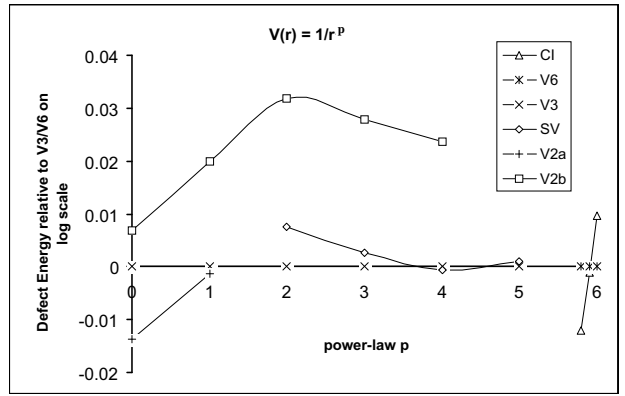}
\smallskip
\caption{Defect energies for $V(r) = 1/r^p$, $n = 5$, on the log-scale, with 
respect to $V_3$/$V_6$. The $CI$ and $V_6$ cross at $p \approx 5.9$. Lines 
joining the data points are only an aid to the eye.}
\label{gam:diff}
\end{figure}
\end{minipage}
\medskip

Following previous authors, the simulations were performed in an almost square 
(length-to-width ratio $5 : 3\sqrt{3}$) cell containing
$N = 5n \times 6n = 30 n^2$ lattice points with $n =$ 1 -- 5 (rather than a 
more nearly square but bigger rectangle of, say,
$7n \times 8n$ ($7 : 4\sqrt{3}$) which would allow us to sample fewer number of 
system sizes $n$ with a given computational limit on $N$). Fig.~\ref{defects} 
corresponds to $n=3$.

A defect is introduced by adding or removing a particle, and then allowing the 
resulting configuration to relax. The difference between the energies of the 
relaxed defect configuration and the perfect lattice configuration gives the 
energy of the defect. There are two modifications to this simple calculation. 
We want the defect energy corresponding to the physical conditions of constant 
chemical potential or line density, so we rescale the cell dimensions (by 
changing the lattice constant $a_0$) after inserting the defect to restore the 
system to its original density (following Ref.~\cite{CockEl:defect}). Moreover, 
since we would ideally like to study an infinite system, the large, but finite 
cell containing $30 n^2$ particles is assumed to be repeated in all directions, 
so that we are effectively dealing with a periodic array of defects, or, an 
infinite lattice in the absence of a defect. The periodic boundary conditions 
maintain the average line density  during the relaxation process. However, now 
the energy per cell also includes the energy of interaction of a defect with 
all its periodic images. As discussed earlier, this energy is finite, and by 
extrapolating its dependence on cell size $n$, i.e., inter-defect separation 
($\approx 5n$), to large $n$, the energy of an isolated defect can be 
extracted~\cite{Frey:defect,CockEl:defect}.

For short-ranged interactions, the energy calculation can be simplified. We 
introduce a cut-off interaction radius $r_c$ where the interaction falls to a 
small fraction of its nearest-neighbour value. The interaction with the 
particles outside can be approximately accounted for by assuming a uniform 
density outside and integrating over it. The radius $r_c$ is chosen to make 
this correction small compared to the total energy, say, less than $10^{-3}$ of 
it. Interactions within the shell are calculated explicitly. As long as
$r_c < L/2$, $L$ being the cell width, this short-range method should be very 
accurate.

For long-ranged interactions such as $\ln{r}$, $1/r$, or $1/r^2$, the above 
method breaks down, and we must resort to the Ewald summation 
technique~\cite{Rosenfeld:Ewald,Heyes:Ewald} which yields an effective 
two-particle interaction that includes the interaction of one particle with all 
the periodic images of the other. This effective potential consists of a real 
space sum (corresponding to a screened interaction) and a reciprocal space sum 
(corresponding to the screening charge). The division between the two is 
controlled by an Ewald parameter, and by a judicious choice of its value, the 
interaction can be made sufficiently short-ranged for both sums. We then employ 
cut-offs in both spaces, with values determined by the desired precision (see 
Appendix C for details).


\noindent
\begin{minipage}{\textwidth}
\begin{table}
\begin{tabular}{|l||c|c|c|c|c|c|}
$\kappa a$ &		$I_3$ &			$SV$ &			$V_{2a}$
&			$V_3$ &			$V_{2b}$ &			$V_6$\\
\hline\hline
0 &	.073016802 &		$V_{2a}$ &		.107018876 &	.108206944 &
.109320135 &		$V_3$\\
1 &	.066331581 &	.096728537 &	.096661116 &	.097578530 &
	.099169907 &		$V_3$\\
2 &	.050588818 &	.072306827 &	.072341149 &	.072594220 &
	.073852944 &		$V_3$\\
3 &	.033575192 &	.046095915 &		$SV$ &		.046131759 &
	.047174061 &		$V_3$\\
4 &	.020037313 &	.025980648 &		$SV$ &		.025962421 &
	.026641900 &		$V_3$\\
\hline\hline
4 &	.020036\hspace*{\fill} &	.025980\hspace*{\fill} &		$SV$ &

		.025961\hspace*{\fill} &	.026641\hspace*{\fill} &		$V_3$\\
5 &	.0110170\hspace*{\fill} &	.0133112\hspace*{\fill} &		$SV$ 
&		.0133146\hspace*{\fill} &	.0136217\hspace*{\fill} &		$V_3$\\
5.1 &	.010338333 &	.012397139 &		$SV$ &		.012400742 &
.012674362 &		$V_3$\\
5.2 &	.009695442 &	.011537972 &		$SV$ &		.011541059 &
		$V_3$ &			$V_3$\\
5.3 &	.009087036 &	.010731274 &		$SV$ &		.010733113 &
		$V_3$ &			$V_3$\\
5.4 &	.008511788 &	.009974612 &		$SV$ &		.009974441 &
		$V_3$ &			$V_3$\\
5.5 &	.007968369 &	.009265581 &		$SV$ &		.009262603 &
		$V_3$ &			$V_3$\\
5.6 &	.007455456 &	.008601808 &		$SV$ &		.008595187 &
		$V_3$ &			$V_3$\\
5.7 &	.006971737 &	.007980968 &		$SV$ &		.007969812 &
		$V_3$ &			$V_3$\\
5.8 &	.006515917 &	.007400791 &		$V_3$ &		.007384121 &		$V_3$ &			$V_3$\\
5.9 &	.006086722 &		$V_6$ &			$V_6$ &
			$V_6$ &			$V_6$ &		.006835768\\
6 &	.005682901 &		$V_6$ &			$V_6$ &
			$V_6$ &			$V_6$ &		.006322377\\
7 &	.002788486 &		$V_6$ &			$V_6$ &
			$V_6$ &			$V_6$ &		.002771295\\
\end{tabular}
\caption{Defect energies for $V(r) = K_0(\kappa r)$; $a_0=1$; system size 
$n = 4$ ($N = 480$). The upper part corresponds to the Ewald Sum method for 
long-range interactions, the lower part to a simple cut-off method for 
short-range interactions. The centered interstitial and the symmetric vacancy 
cross at $\kappa a \approx 6.9$. Entries such as ``$V_{2a}$'', ``SV'', 
``$V_3$'', ``$V_6$'' indicate an instability to a lower energy defect.}
\label{tbl:bes}
\end{table}
\end{minipage}


\noindent
\begin{minipage}{\textwidth}
\begin{table}
\begin{tabular}{|l||c|c|c|c|c|c|}
\ $p$&		$I_3$&			$SV$&			$V_{2a}$&
$V_3$&			$V_{2b}$&			$V_6$\\
\hline\hline
\ 0&	\ 0.073061685&		$V_{2a}$&		0.106775085&	0.108253779&
	0.108994418&		$V_3$\\
\ 1&	\ 0.146421440&		$V_{2a}$&		0.209046876&	0.209331872&
	0.213568209&		$V_3$\\
\ 2&	\ 0.487928019&	0.677444176&		$SV$&		0.672359275&
	0.694143882&		$V_3$\\	
\ 3&	\ 1.08543992\hspace*{\fill}&		1.39071722\hspace*{\fill}&
			$SV$&		1.38704618\hspace*{\fill}&		1.42628053\hspace*{\fill}&			$V_3$\\
\ 4&	\ 1.99663790\hspace*{\fill}&		2.37494467\hspace*{\fill}&
			$SV$&		2.37649196\hspace*{\fill}&		2.43341170\hspace*{\fill}&			$V_3$\\
\ 5&	\ 3.2620983\hspace*{\fill}&		3.5889518\hspace*{\fill}&
			$SV$&		3.5851010\hspace*{\fill}&			$V_3$&
			$V_3$\\
\ 5.8&	\ 4.5498400\hspace*{\fill}&			$V_6$&
			$V_6$&			$V_6$&			$V_6$&		\ 4.6053332\\
\ 5.9&	\ 4.7286554\hspace*{\fill}&			$V_6$&
			$V_6$&			$V_6$&			$V_6$&
		\ 4.7341340\\
\ 6&	\ 4.9114956\hspace*{\fill}&			$V_6$&			$V_6$&
			$V_6$&			$V_6$&		\ 4..8637723\\
\ 7&	\ 6.9642383\hspace*{\fill}&			$V_6$&			$V_6$&
			$V_6$&			$V_6$&		\ 6.1999848\\
\ 8&	\ 9.4317462\hspace*{\fill}&			$V_6$&			$V_6$&
			$V_6$&			$V_6$&		\ 7.5920876\\
\ 9&	12.319586\hspace*{\fill}&			$V_6$&			$V_6$&
			$V_6$&			$V_6$&		\ 9.0220754\\
10&	15.629229\hspace*{\fill}&			$V_6$&			$V_6$&
			$V_6$&			$V_6$&		10.477581\hspace*{\fill}\\
11&	19.359421\hspace*{\fill}&			$V_6$&			$V_6$&
			$V_6$&			$V_6$&		11.950259\hspace*{\fill}\\
12&	23.495660\hspace*{\fill}&			$V_6$&			$V_6$&
			$V_6$&			$V_6$&		13.434556\hspace*{\fill}\\
\end{tabular}
\caption{Defect energies for $V(r) = 1/r^p$; $a_0=1$; system size $n = 5$ 
($N = 750$). The Ewald sum technique was used to calculate the energies. The 
centered interstitial and the symmetric vacancy cross at $p \approx 5.9$. 
Entries such as ``$V_3$'' and ``$V_6$'' indicate an instability to a lower 
energy defect.}
\label{tbl:gam}
\end{table}
\end{minipage}

To find the minimum of the interaction energy as a function of the 
configuration of N particles, we use the conjugate-gradient 
method~\cite{Num-Rec}. The forces are also needed for this method, and are 
easily derived from the energy and conveniently calculated along with it.

The results for $n = 4$ ($480$ particles) for $V_{\kappa a}$ and for $n = 5$ 
for $V_p$ ($750$ particles) are shown in Tables~\ref{tbl:bes} and~\ref{tbl:gam}
and Figs.~\ref{bes:ek} and~\ref{gam:ep}. ($n = 5$ was computationally 
prohibitive for the long-ranged regime with $\kappa a_0 > 0$). Note that, for 
the screened Bessel-function interaction, we find that calculations optimized 
for the long- and short-ranged regimes agree to within $1$ part in $20,000$ at 
$\kappa a_0 = 4$.

Moreover, we find that the interaction of a defect with all 
its periodic images is repulsive for defects with (even) two- and six-fold 
symmetry, and attractive for (odd) three-fold symmetry, consistent with 
Ref.~\cite{Frey:defect}. As discussed in
Refs.~\cite{Frey:defect} and~\cite{CockEl:defect}, the true asymptotic form of 
the power law defect interaction probably isn't reached for the distance scales 
$r \sim 20-30$ lattice spacings studied here.


\section{Conclusions}
We have studied factors contributing to the wandering of a vacancy or 
interstitial string defect in a hexagonal columnar crystal. A gas of such 
strings in the crystalline phase, interacting via short-range potentials, can 
proliferate via continuous or first-order transitions when the corresponding 
defect chemical potential changes sign, leading to a supersolid phase. The 
transition can be modified by the presence of vacancy-interstitial loops, 
especially in a system of finite thickness. We have also numerically calculated 
defect line tensions for two families of line interactions which interpolate 
between long- and short-ranged interaction potentials. In each case, we 
determine the point where interstitial and vacancy defects exchange stability. 
A complete accounting requires consideration of a variety of nearly degenerate 
vacancy configurations. At finite temperatures, the small energy differences 
between diffferent species will further lower the free energy of the vacancy 
through a gain in fluctuation entropy. The interstitial itself can fluctuate 
between the centered and edge configurations. The point where vacancies and 
interstitials exchange stability will shift at finite temperatures due to 
entropic effects of this kind. In the context of long-range potential 
calculations, we show in an Appendix how to extend the Ewald summation to the 
modified Bessel function potential $K_0(x)$.

\acknowledgements

This research was supported by the National Science Foundation, in part by 
the MRSEC Program through Grant No. DMR-9400396 and through
Grant No. DMR-9714725.

\appendix

\section{Calculation of energy of distortion due to a defect string}
\label{bend}

As described in Ref.~\cite{SelB:defect}, minimization of the free energy 
(\ref{Ftot}) with the constraint (\ref{constraint}) yields the following 
equation for
${\mathbf u}({\mathbf r}_\perp,z)$:
\begin{equation}
\lambda_L^2 \partial_z^4 {\mathbf u} -
{\mathbf \nabla}_\perp ({\mathbf \nabla}_\perp \cdot {\mathbf u}) =
\frac{1}{\rho_0} {\mathbf \nabla}_\perp \delta({\mathbf r}_\perp-{\mathbf r}_d)
\end{equation}
${\mathbf r}_d$ being the in-plane location of the defect string (assumed 
straight for now).
Upon assuming a solution of the form
${\mathbf u} = -\frac{1}{\rho_0} {\mathbf \nabla}_\perp \psi$, we have the 
scalar equation
\begin{equation}
\label{psi}
(-\lambda_L^2 \partial_z^4 + \nabla_\perp^2) \psi =
\delta({\mathbf r}_\perp-{\mathbf r}_d)
\end{equation}
For the straight string, the solution is
\begin{equation}
\psi({\mathbf r}_\perp,z) \propto \ln{|{\mathbf r}_\perp-{\mathbf r}_d|},
\quad\mbox{or,}\quad
{\mathbf u}({\mathbf r}_\perp,z) \propto
\frac{{\mathbf r}_\perp-{\mathbf r}_d}
{\left|{\mathbf r}_\perp-{\mathbf r}_d\right|^2} ,
\end{equation}
with proportionality constant $\sim a_0^2$.

Now consider a wandering string with a dilute concentration of kinks, described 
on average by ${\mathbf r}_d(z)$ (see Fig.~\ref{string}). Upon inserting this 
$z$-dependence into the right hand side of Eq.~(\ref{psi}), we see that the 
resulting $\psi$ inherits the fluctuations of ${\mathbf r}_d(z)$.
If $l_z$ represents the smallest wavelength in ${\mathbf r}_d(z)$, the two 
terms on the LHS of Eq.~(\ref{psi}) compare as $\lambda_L^2/l_z^4$ vs. 
$1/a_0^2$, or as $l^*$ vs. $l_z$ where $l^* = \sqrt{\lambda_L a_0}\;$ is of the 
order of the kink length. Since the meandering of the defect string occurs on a 
length scale much larger than the kink size, the first term should be 
negligible compared to the second, and we can set
\begin{equation}
\psi({\mathbf r}_\perp,z) \propto \ln{|{\mathbf r}_\perp-{\mathbf r}_d(z)|}
\end{equation}
as a reasonable approximation.

The elastic energy of a defect of length $L$ can now be written as
\begin{equation}
E_{defect} = \tau_z L +
\varepsilon_k \int\!\! \frac{dz}{a_0} \left|\frac{d{\mathbf r}_d(z)}{dz}\right|
+ \frac{1}{2}
\int'\!\! d^{3}r K_3 \left(\frac{\partial^2{\mathbf u}}{\partial z^2}\right)^2
\end{equation}
representing contributions from line tension, kinks, and the bending energy of 
the distorted crystal (zero for a straight string).
The primed integral here excludes the core of the string: a region of radius 
$\sim a_0$ around it. It can easily be evaluated for
${\mathbf u}({\mathbf r},z) =
{\mathbf u}_d({\mathbf r}_\perp-{\mathbf r}_d(z),z)$ and reduces to the form in 
Eq.~(\ref{defE4}), accurate up to fourth order in the derivatives.
The second term, on the other hand, leads to the term
$(g/2)\int\!\! dz |d{\mathbf r}_d/dz|^2$ in Eq.~(\ref{defEg}). For long 
wavelengths, the additional contribution from the third term is irrelevant in 
comparison, being of higher order in the derivatives. The length scale at which 
it becomes important is obtained by balancing the two terms:
$K_3/l_z^4 \sim T/D/l_z^2$, or, $l_z \sim \sqrt{K_3 D/T}$.

\section{Renormalization of D by defect-phonon coupling}
\label{DR}

\begin{minipage}{2.3in}
\begin{figure}
\centering
\leavevmode
\epsfxsize=1.5in
\epsfbox{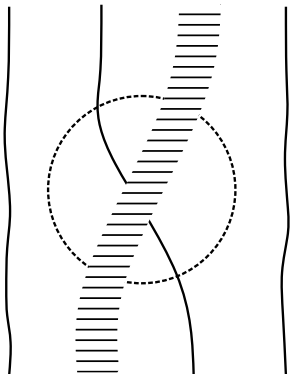}
\smallskip
\caption{Illustration of the coupling between a defect string and the lattice 
distortion. In this case, the change in the position of the vacancy string 
(thick dashed curve) is equal and opposite to the change in the phonon 
displacement field.}
\label{couple}
\end{figure}
\end{minipage}
\hfill
\setlength{\textw}{\textwidth}
\addtolength{\textw}{-2.6in}
\begin{minipage}{\textw}
In Section~\ref{single} we described the wandering of a defect line along the 
$\hat{\mathbf z}$-axis by a ``diffusion'' constant $D = a_0^2 n_k/2$, 
corresponding to an effective Hamiltonian (Eq.~(\ref{defEg}))
$H_{defect} = (g/2) \int\!\! dz |d{\mathbf r}_d/dz|^2$, ${\mathbf r}_d(z)$ 
describing the in-plane position of the defect string, with $g=T/D$. To 
incorporate the effect of lattice fluctuations on the diffusion of the defect 
string, we modify $H_{defect}$ to
\begin{equation}
H_{defect\mbox{\small\textit{-}}phonon} = \frac{g}{2} \int\!\! dz
\left|\pm \frac{d{\mathbf r}_d}{dz} + {\mathbf t}({\mathbf r}_d,z)\right|^2
\end{equation}
where the expression in brackets now represents the deviation of the 
vacancy/interstitial string with respect to the local director
\begin{equation}
{\mathbf t} \equiv \frac{\partial \mathbf{u}}{\partial z}
\end{equation}
Fig.~\ref{couple} illustrates the case of a vacancy string, which we shall 
assume for the remainder of this appendix.
\end{minipage}
\smallskip

It is easy to derive the diffusion equation for the partition function 
${\mathcal Z}({\mathbf r}_d, {\mathbf r}_0; z, 0)$ corresponding to the above 
Hamiltonian (the \{${\mathbf u}({\mathbf r}_\perp,z)$\}-dependence in 
${\mathcal Z}$ has been omitted for convenience):
\begin{equation}
\partial_z {\mathcal Z} - ({\mathbf t} \cdot {\mathbf \nabla}_\perp){\mathcal Z}
= D \nabla_\perp^2 {\mathcal Z}
\end{equation}
${\mathcal Z}$ represents the probability density for the defect position; 
$-{\mathbf t}$ is the ``convective velocity'' for this density. It can also be 
thought of as an (imaginary) vector potential acting on a particle of mass $g$ 
in two dimensions, with $z$ the time-like coordinate.

\begin{figure}
\centering
\leavevmode
\epsfxsize=3in
\epsfbox{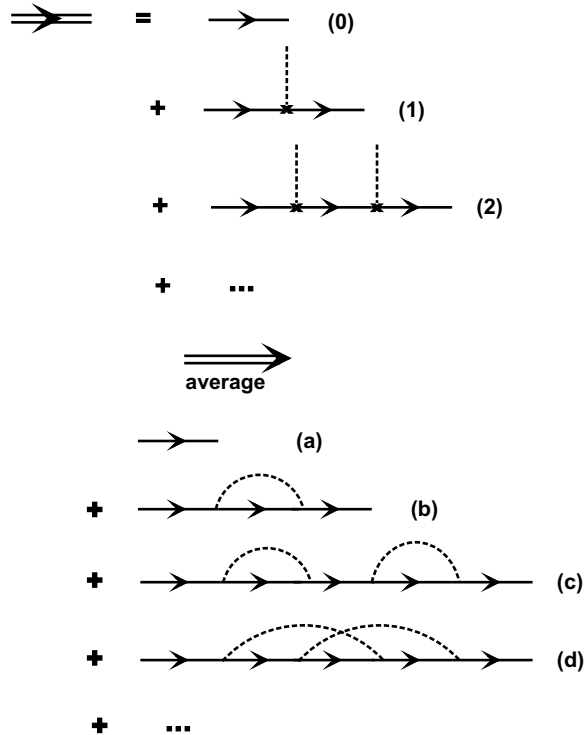}
\medskip
\caption{Diagrammatic representation of the series expansion of the propagator 
for the defect probability density. The average is over the phonon degrees of 
freedom.}
\label{feynman}
\end{figure}

Defining the propagator $G({\mathbf r}_\perp,z) =
{\mathcal Z}({\mathbf r}_\perp,z) \theta(z)$, $\theta(z)$ being the step 
function, $G$ obeys
\begin{equation}
\label{Gr}
(\partial_z - D \nabla_\perp^2) G({\mathbf r}_\perp,z) =
\delta^{(2)}({\mathbf r}_\perp) \delta(z) +
{\mathbf t} \cdot {\mathbf \nabla}_\perp G
\end{equation}
The bare propagator $G_0$ corresponds to ignoring the convective influence of 
the medium. Thus, $G_0$ satisfies
\begin{equation}
(\partial_z - D \nabla_\perp^2) G_0({\mathbf r}_\perp,z) =
\delta^{(3)}({\mathbf r})
\end{equation}
Fourier-transforming ${\mathbf r}_\perp \rightarrow {\mathbf k}$ (space-like) 
and $z \rightarrow \omega$ (time-like),
\begin{equation}
G_0({\mathbf k},\omega) = \left(-i \omega + D {\mathbf k}^2\right)^{-1}
\end{equation}
The renormalized diffusion coefficient $D_R$ will be calculated from the 
average of $G$ over the phonon degrees of freedom using the definition
\begin{equation}
G({\mathbf k},\omega)^{-1} = -i \omega + D_R k^2
\end{equation}
in the limit $|{\mathbf k}|, \omega \rightarrow 0$. Upon denoting
$k \equiv ({\mathbf k},\omega)$, Eq.~(\ref{Gr}) becomes
\begin{equation}
\label{Gk}
G_0^{-1}(k) G(k) = 1 + \int_{k'} i {\mathbf k}' \cdot {\mathbf t}(k-k') G(k')
\end{equation}
The symbol $\int_k$ denotes $\int d^{3}k/(2\pi)^3$.
Eq.~(\ref{Gk}) can be expanded in a perturbation series:
\begin{eqnarray}
G(k) &=& G_0(k) +
G_0(k) \int_{k'} i {\mathbf k}' \cdot {\mathbf t}(k-k') G_0(k')	\nonumber\\
& & \mbox{  } + G_0(k) \int_{k'} i {\mathbf k}' \cdot {\mathbf t}(k-k') G_0(k')
\int_{k''} i {\mathbf k}'' \cdot {\mathbf t}(k'-k'') G_0(k'') + \ldots
\end{eqnarray}
To calculate the thermal averages of products of
${\mathbf t} = -i \omega {\mathbf u}$, we need
\begin{eqnarray}
\langle u_\alpha(k) \rangle &=& 0,	\nonumber\\
\langle u_\alpha(k) u_\beta(k') \rangle &=&
\left[ S_L(k) {\mathcal P}^L_{\alpha \beta}(k) +
S_T(k) {\mathcal P}^T_{\alpha \beta}(k) \right] \delta^{(3)}(k-k')	\\
& \equiv & S_{\alpha \beta}(k) \delta^{(3)}(k-k')
\end{eqnarray}
where the correlation functions parallel (L) and perpendicular (T) to
${\mathbf k}$ are
\begin{equation}
S_{L/T}(k) = \frac{T}{K_3 \omega^2 + c_{11/66} k^2}
\end{equation}
and the projection operators are ${\mathcal P}^L_{\alpha \beta}(k) =
k_\alpha k_\beta/k^2$ and ${\mathcal P}^T_{\alpha \beta}(k) =
\delta_{\alpha \beta} - {\mathcal P}^L_{\alpha \beta}(k)$. Therefore
\begin{equation}
\label{series}
\langle G(k) \rangle = G_0(k) -
G_0(k) \left[ \int_{k'} k_\alpha k'_\beta \omega^2 S_{\alpha \beta}(k-k')
G_0(k') \right] G_0(k) + \ldots
\end{equation}
Diagrammatically, this series is represented in Fig.~\ref{feynman}.

All diagrams of type (b), (c) reducible to the one-loop diagram (b) can be 
summed to give $\langle G(k) \rangle^{-1} \approx G_0(k)^{-1} + V(k)$ where 
$V(k)$ is the term in square brackets in Eq.~(\ref{series}). Then,
$D_R = D + \delta D_R$ where
\begin{equation}
\delta D_R \approx \lim_{k \rightarrow 0} \frac{1}{2} \nabla_{\mathbf k}^2 V(k)
\end{equation}
With the assumption $\lambda_{L/T} D^2 a_0^{-3} \ll 1$, in other words, kink 
size $l^* \ll l_k$, the kink separation, $\delta D_R$ can be evaluated to give
\begin{eqnarray}
\frac{\delta D_R}{D} \approx \frac{T}{K_3^{1/4}} \left(\frac{1}{c_{11}^{3/4}} +
\frac{1}{c_{66}^{3/4}}\right) \frac{\Lambda^{5/2}}{20 \sqrt{2} \pi}
\end{eqnarray}
where we have imposed a cutoff by replacing the hexagonal Brillouin zone by a 
circle of radius $\Lambda \equiv \frac{4 \pi}{\sqrt{3} a_0}$ which has the same 
area.
If $c_{66} \ll c_{11}$, the fluctuations are mostly transverse, and
\begin{equation}
\frac{\delta D_R}{D} \approx
\frac{\langle |{\mathbf u}|^2 \rangle}{a_0^2} \frac{4\pi}{5\sqrt{3}}
\end{equation}
Since $\frac{\langle |{\mathbf u}|^2 \rangle}{a_0^2} \lesssim c_L^2$ where the 
Lindemann ratio for hexagonal columnar crystalline lattices is empirically 
known to be $c_L^2 \simeq 1/50$~\cite{cl}, we find $\frac{\delta D_R}{D} < 3\%$.

\section{Interaction between defects with $n$-fold symmetry}
\label{n-def}

We assume the defect string to be straight, so that only the planar elastic 
deformation energy, ${\mathcal F}_{crystal}$ (Eq.~\ref{Fxtal}) is relevant. In 
the continuum model discussed, this energy is isotropic in the strains $u_{ij}$.
The stress $\sigma_{ij} = \delta {\mathcal F}_{crystal}/\delta u_{ij}$ can be 
expressed in terms of a biharmonic stress function $\chi$\cite{LL}: 
$\nabla_\perp^2 \nabla_\perp^2 \chi = 0$, as
$\sigma_{ij} = \epsilon_{ik} \epsilon_{jl} \partial_k \partial_l \chi$
($\epsilon_{ij}$ is the two-dimensional anti-symmetric tensor,
$\epsilon_{12}=1$).

For a dislocation with Burger's vector ${\mathbf b}$,
$\chi = -K {\mathbf b} \times {\mathbf r}_\perp \ln{r_\perp}$ where
$K=\mu (\lambda+\mu)/\pi (\lambda+2\mu)$ in terms of the Lam\'{e} coefficients.

We construct $\chi$ for an $n$-fold symmetric vacancy/interstitial by treating 
it as a superposition of $n$ dislocations (bound) symmetrically placed a 
distance $d \approx a_0$ apart (such that the volume of the defect is
$\Omega \sim n d a_0$), with Burger's vectors separated by $2\pi/n$ in 
orientation. The resulting stress function has a form satisfying
\begin{equation}
\nabla_\perp^2 \chi = K \Omega \left(2\pi \delta^{(2)}({\mathbf r}_\perp) -
\frac{1}{d^2}\sum_{k=1}^\infty
a_k \frac{\cos{k n \theta}}{(r_\perp/d)^{kn}}\right)
\end{equation}
where $a_k$ are coefficients of ${\mathcal O}(1)$.

The interaction energy of two such defects, located at ${\mathbf r_1}$ and
${\mathbf r_2}$ respectively, can be written in terms of their stress functions 
as
\begin{equation}
{\mathcal U}_{12}({\mathbf r_{12}}\equiv{\mathbf r_2}-{\mathbf r_1}) =
\frac{1}{4\pi K} \int d^2 r_\perp
\nabla_\perp^2 \chi({\mathbf r}_\perp-{\mathbf r_1})
\nabla_\perp^2 \chi({\mathbf r}_\perp-{\mathbf r_2}) .
\end{equation}
For $r_{12}/d \gg 1$, the leading term in the interaction comes from the 
convolution of the $\delta$-function with the $k=1$ term (in other words, this 
is the cost of the volume change produced by one defect in the stress field of 
the other), therefore it is of the form $\cos{n\theta_{12}}/{r_{12}}^n$.

Specifically, we find for vacancies (the sign is reversed for interstitials):
\begin{center}
\begin{tabular}{|c|c|c|}
\hline
$n$&	relative  orientation&	${\mathcal U}({\mathbf r}_\perp)$ in units of 
$K\Omega/d^2$\\
\hline
2&	parallel&		$\frac{-\cos{2\theta}+\cos{4\theta}}{(r/d)^2}$\\
2&	perpendicular&	$\frac{-2\cos{2\theta}+\cos{4\theta}}{(r/d)^2}$\\
3&	parallel&		$\frac{-3\cos{6\theta}}{(r/d)^4}$\\
3&	anti-parallel&
$\frac{-2\cos{3\theta}}{(r/d)^3}+\frac{3\cos{6\theta}}{(r/d)^4}$\\
6&	&			$\frac{-2\cos{6\theta}}{(r/d)^6}$\\
\hline
\end{tabular}
\end{center}

\section{The Ewald Sum for $V(r) =$ $K_0(\kappa r)$ and $1/r^p$}
\label{Ewald}

Let $\phi(r)$ be the two-body interaction potential between charges $q_i$ in 
the given system so that the total interaction energy is
\begin{equation}
U = \frac{1}{2} \mathop{\sum \sum}_{i \neq j} q_i q_j \phi(r_{ij})
\end{equation}
The simulated system consists of N particles in a cell repeated to generate an 
infinite system. Then, the energy per cell can be written as 
\begin{equation}
U = U_{0_\alpha} +
\frac{1}{2} \mathop{\sum \sum}_{i \neq j}^N q_i q_j v_\alpha(r_{ij})
\end{equation}
where~\cite{Rosenfeld:Ewald}
\begin{mathletters}
\begin{equation}
v_\alpha(r_{ij}) =
\sum_{{\mathbf n}} \left[\phi({\mathbf r}_{ij}+{\mathbf n})-
\psi_\alpha({\mathbf r}_{ij}+{\mathbf n})\right] +
\frac{1}{A} \sum_{{\mathbf G}}
\tilde{\psi_\alpha}({\mathbf G}) e^{i {\mathbf G}\cdot{\mathbf r}_{ij}} 
\end{equation}
\begin{equation}
U_{0_\alpha} = \frac{1}{2} \left(\sum_i {q_i}^2\right) \lim_{{\mathbf r}
\rightarrow 0} \left[v_\alpha({\mathbf r})-\phi({\mathbf r})\right]
\end{equation}
\end{mathletters}
Here the sum over ${\mathbf n}$ consists of all real space lattice vectors of 
the lattice generated by a cell of area $A$, and the sum over ${\mathbf G}$ 
goes over the corresponding reciprocal lattice vectors. The lattice is 
rectangular (almost square) in our case, which makes it easy to list these 
vectors. $\psi_\alpha$ is the long-range part of the interaction $\phi$, so 
that $\phi-\psi_\alpha$ is a screened, short-ranged interaction. 
$\tilde{\psi_\alpha}$ is the Fourier transform. The amount of screening is 
controlled by the Ewald parameter $\alpha$. If we take $\alpha$ to be large 
enough so that the real space sum can be truncated at $r_c = L/2$ within the 
desired precision, then we can drop the sum over
${\mathbf n} \neq 0$~\cite{AdamsMcD:Ewald}. However, this means including more 
short-range components in the screening charge distribution, so that it spreads 
to higher reciprocal vectors. The cutoff in reciprocal space is again 
determined by the precision required.

Since we shall be considering $N$ particles, each with $q_i = 1$, we have 
$\sum_i {q_i}^2 = N$ and $(\sum_i q_i)^2 = N^2$. Rearranging the sums in $U$ 
and noting that $N/A \equiv \rho$ which is constant throughout the calculation, 
we can rewrite $U$ as
\begin{equation}
U = U_{ref} + U_{int}
\end{equation}
where~\cite{Rosenfeld:Ewald,AdamsMcD:Ewald}
\begin{mathletters}
\begin{equation}
U_{int} \approx
\mathop{\sum \sum}_{i < j}^N \left[\phi({\mathbf r}_{ij})-
\psi_\alpha({\mathbf r}_{ij})\right] +
\frac{1}{2 A} \sum_{{\mathbf G}} \tilde{\psi_\alpha}({\mathbf G})
\left[\left(\sum_i \cos{{\mathbf G}\cdot{\mathbf r}_i}\right)^2 +
\left(\sum_i \sin{{\mathbf G}\cdot{\mathbf r}_i}\right)^2\right]
\end{equation}
\begin{equation}
U_{ref} \approx
-\frac{N}{2} \lim_{{\mathbf r} \rightarrow 0} \psi_\alpha({\mathbf r}) +
\frac{N}{2} \rho
\lim_{{\mathbf k} \rightarrow 0} \tilde{\psi_\alpha}({\mathbf k})
\end{equation}
\end{mathletters}
Note that $U_{ref}$ is explicitly proportional to $N$ in this form. This is the 
form we use in our calculations. For interactions whose long-range integral 
diverges (such as $1/r^p$ with $p \leq 2$), a uniformly spread background of 
equal and opposite charge is assumed, so that the second term in $U_{ref}$ 
should contain
\begin{equation}
\lim_{{\mathbf k} \rightarrow 0}
\left[\tilde{\psi_\alpha}({\mathbf k})-\tilde{\phi}({\mathbf k})\right] .
\end{equation}

We can now proceed to the special potentials we are interested in.
For the power-law potential
\begin{equation}
\phi(r) = 1/r^p \equiv
\frac{1}{\Gamma(p/2)} \int_0^\infty dt \; t^{p/2-1} e^{-t r^2},
\end{equation}
we take~\cite{Nijboer:gam}
\begin{equation}
\psi_\alpha(r) =
\frac{1}{\Gamma(p/2)} \int_{\alpha^2}^\infty dt \; t^{p/2-1} e^{-t r^2} \equiv
\frac{1}{r^p} \frac{\Gamma\left(p/2,(\alpha r)^2\right)}{\Gamma(p/2)},
\end{equation}
so that the screened interaction is
\begin{equation}
\phi(r)-\psi_\alpha(r) =
\frac{1}{r^p} \frac{\gamma\left(p/2,(\alpha r)^2\right)}{\Gamma(p/2)}. 
\end{equation}
($\Gamma$ and $\gamma$ are complementary incomplete Gamma
functions~\cite{Math-Func}.)

The Fourier transform is ($d = 2$):
\begin{equation}
\tilde{\psi_\alpha}(k) =
\frac{\pi^{d/2}}{\Gamma(p/2)}
\left(\frac{2}{k}\right)^{d-p}
\Gamma\left(\frac{d-p}{2},\left(\frac{k}{2 \alpha}\right)^2\right)
\end{equation}
Also,
\begin{equation}
\lim_{r \rightarrow 0} \psi_\alpha(r) = \frac{\alpha^p}{\Gamma(p/2+1)},
\end{equation}
\begin{equation}
\lim_{k \rightarrow 0} \tilde{\psi_\alpha}(k) =
\frac{2}{p-d} \frac{\pi^{d/2} \alpha^{p-d}}{\Gamma(p/2)}, \quad p > d
\end{equation}
For $p < d$, the above expression corresponds to
\( \lim_{k \rightarrow 0} \left[\tilde{\psi_\alpha}(k)-\tilde{\phi}(k)\right]\).
For $p = d$, both forms would diverge, however, they would be independent of 
$\alpha$, and since the defect energy is a difference of energies, this term 
would cancel out.

The force on particle $j$, ${\mathbf f}_j \equiv -{\mathbf \nabla}_j U$, can 
also be written as a sum of real space and reciprocal space contributions:
\begin{mathletters}
\label{f}
\begin{equation}
{\mathbf f}_j = \sum_{i \neq j} {\mathbf f}_{ij}^R + {\mathbf f}_j^G
\end{equation}
where
\begin{equation}
{\mathbf f}_{ij}^R = 2
\frac{\Gamma\left(p/2+1,(\alpha r_{ij})^2\right)}{r_{ij}^{p+2}} {\mathbf r_{ij}}
\end{equation}
is the sum of forces on particle $j$ due to particle $i$ and all its images, and
\begin{equation}
{\mathbf f}_j^G =
\frac{1}{A} \sum_{{\mathbf G} \neq {\mathbf 0}} \tilde{\psi_\alpha}(G)
\left[\left(\sum_i
\cos{{\mathbf G}\cdot{\mathbf r}_i}\right) \sin{{\mathbf G}\cdot{\mathbf r}_j} -
\left(\sum_i
\sin{{\mathbf G}\cdot{\mathbf r}_i}\right) \cos{{\mathbf G}\cdot{\mathbf r}_j}
\right] {\mathbf G}
\end{equation}
is the sum of forces on particle $j$ due to all images of itself.
\end{mathletters}

The co-ordinates ${\mathbf r}$ here are normalized such that $a_0 = 1$. When we 
change $N$ to $N_d = N \pm 1$ and rescale $a_0$ to $a = \sqrt{N_d/N}$ after 
inserting a defect, we chose to keep ${\mathbf r}$ normalized with respect to 
$a$, so that it picks up a factor of $a$. If we also scale $\alpha$ by $1/a$, 
the product $\alpha r$ remains unchanged (as does ${\mathbf G}\cdot{\mathbf r}$), so that we can keep using the original values of $\alpha$ and ${\mathbf G}$ in
$U_{int}$ \& ${\mathbf f}$, and scale the result by $1/a^p$ in the end. In 
$U_{ref}$ we have to use the scaled value of $\alpha$ along with $N_d$, and 
subtract the energy of the perfect lattice scaled by $N_d/N$.

On the other hand, if we do not scale $\alpha$, $U_{ref}$ cancels out, but 
$\alpha$ has to be replaced by $\alpha a$ in $U_{int}$ \& ${\mathbf f}$.

For the modified Bessel function interaction
\begin{equation}
\phi(r) = K_0(\kappa r),
\end{equation}
(where $\kappa$ represents $\kappa a$ because of the normalization of $r$),
the $\kappa = 0$ case, corresponding to $\phi(r) \sim -\ln{r}$, has been 
treated by Frey \textit{et al.}~\cite{Frey:defect}. To extend this to
$\kappa > 0$, we perform an expansion similar to that of
Mokross \& Silva~\cite{MokSil:series} for a Yukawa potential. Writing the 
potential in integral form,
\begin{equation}
\phi(r) = K_0(\kappa r) \equiv
\frac{1}{2} \int_0^\infty \frac{dt}{t} e^{-t} e^{-\frac{(\kappa r)^2}{4 t}}
\end{equation}
We choose the screened interaction to be
\begin{equation}
\phi(r) - \psi_\alpha(r) = \frac{1}{2}
\int_{(\alpha r)^2}^\infty \frac{dt}{t} e^{-t} e^{-\frac{(\kappa r)^2}{4 t}} =
\frac{1}{2} \int_1^\infty \frac{ds}{s}
e^{-(\alpha r)^2 s} e^{-\left(\frac{\kappa}{2 \alpha}\right)^2 \frac{1}{s}}
\end{equation}
Expanding the exponential within the integral in a Taylor series about
$\kappa = 0$, we get
\begin{equation}
\phi(r) - \psi_\alpha(r) = \frac{1}{2} \sum_{n = 0}^\infty \frac{(-1)^n}{n!}
\left(\frac{\kappa}{2 \alpha}\right)^{2 n} E_{n+1}\left((\alpha r)^2\right)
\end{equation}
where $E_n(x)$ is the exponential integral function
($E_{n+1}(x) = x^n \Gamma(-n,x)$).

For $\kappa = 0$, only the first term, $E_1\left((\alpha r)^2\right)$, is 
non-zero.
For $\kappa > 0$ we have an alternating series, and its convergence has to be 
taken into account in determining the optimum value of $\alpha$ (in addition to 
the required precision and the cell size). For large values of $\kappa$, not 
only are a large number of terms needed in this series to reach the desired 
precision, the optimum value of $\alpha$ is also large due to convergence 
rquirements, increasing the cutoff in reciprocal space, so that the computation 
time increases dramatically. We were able to carry these calculations to 
$\kappa = 4$, where it matched the results from the short-range method to $1$ 
part in $20,000$.

We also need the following quantities (in $d = 2$):
\begin{equation}
\tilde{\psi_\alpha}(k) =
2 \pi \frac{e^{-\frac{\kappa^2 + k^2}{(2 \alpha)^2}}}{\kappa^2 + k^2}
\end{equation}
\begin{equation}
\lim_{k \rightarrow 0} \tilde{\psi_\alpha}(k) =
\frac{2 \pi}{\kappa^2} e^{-\left(\frac{\kappa}{2 \alpha}\right)^2}
\end{equation}
\begin{equation}
\lim_{r \rightarrow 0} \psi_\alpha(r) =
\frac{1}{2} E_1\left(\left(\frac{\kappa}{2 \alpha}\right)^2\right) 
\end{equation}
At $\kappa = 0$ we take
\begin{equation}
\lim_{\kappa \rightarrow 0} \lim_{k \rightarrow 0} \left[\tilde{\psi_\alpha}(k)-
\tilde{\phi}(k)\right] = -\frac{\pi}{2 \alpha^2}
\end{equation}
Also,
\begin{equation}
\lim_{r \rightarrow 0} \psi_\alpha(r) \approx -\gamma/2 + \ln{\alpha} -
 \ln{(\kappa/2)} + {\mathcal O}(\kappa^2)
\end{equation}
The $\ln{\kappa}$ term cancels in the defect energy, so that the limit
 $\kappa \rightarrow 0$ is again well-defined. Similarly, if we did not 
 subtract $\lim_{k \rightarrow 0} \tilde{\phi}(k)$, we would have an extra term 
 $2 \pi/\kappa^2$, which too would cancel out.

The expression for the force is similar to Eqs.~(\ref{f}) where
${\mathbf f}_{ij}^R$ is a series similar to $\phi(r) - \psi_\alpha(r)$, with 
each $E_{n+1}\left((\alpha r)^2\right)$ replaced by
$2 \alpha^2 E_n\left((\alpha r)^2\right) {\mathbf r}$. On scaling, 
$\tilde{\psi_\alpha}(k)$ has to be recalculated because it does not simply 
scale as a power law.

The special functions used here were all calculated to an accuracy of $10^{-16}$
according to routines taken from Ref.~\cite{Num-Rec}. Since the power-law 
calculations were mostly carried out on integral values of $p$, the gamma 
functions were only needed for integral or half-integral orders, in which case 
certain recursion relations can be used~\cite{Comp:gam}. We used the fastest 
method for each order.


\end{document}